\documentclass[preprint,draft,showpacs,amsmath,amsfonts,amssymb,aps,prd]{revtex4}

\begin{document}

\title{Path Integral Treatment of 
 Singular Problems and Bound States:
\\
Quantum Mechanics}
                                                                               
\author{
Horacio E. Camblong$^{a}$
and
Carlos R. Ord\'{o}\~{n}ez$^{b,c}$}

\affiliation{$^a$ 
Department of Physics, University of San Francisco, San
Francisco, CA 94117-1080 
\\
$^b$ Department of Physics, University of Houston, Houston,
TX 77204-5506
\\
$^c$
World Laboratory Center for Pan-American Collaboration in Science and
Technology 
}

\begin{abstract}
A path-integral approach for the computation of quantum-mechanical 
propagators and energy Green's functions is presented.
Its effectiveness is demonstrated through its application to singular interactions, with particular 
emphasis on the inverse square potential---possibly combined with a delta-function interaction.
The emergence of these singular potentials as low-energy nonrelativistic limits of
quantum field theory is highlighted.
Not surprisingly, the analogue of ultraviolet regularization is required
for the interpretation of these singular problems.
\end{abstract}


\maketitle

\section{Introduction}
\label{sec:introduction}

It has been generally recognized
that the treatment of singular problems in quantum mechanics is 
notoriously difficult and plagued with ambiguities~\cite{fra:71}.
A complete solution of a typical quantum-mechanical problem with a singular potential requires 
the analysis of the scattering and bound-state sectors supplemented by
a prescription, which can be conveniently formulated in the field-theory language 
of regularization and renormalization. 
Such a prescription
establishes the necessary connection between the physics of
a particular system and the corresponding renormalization of an ill-defined problem.
As this paper demonstrates,
much work is still needed in formulating and interpreting
singular problems in quantum physics,
 particularly within the path-integral framework.

It is also generally accepted that 
our current
understanding 
of bound states in quantum field theory---including relativistic effects and 
radiative corrections---is still unsatisfactory~\cite{wei:95,jac:private}.
Conceptually, the source of this deficiency lies in the apparent
contradiction between the usual perturbative framework of quantum field
theory and the intrinsic nonperturbative nature 
of bound states, which manifest themselves
as poles of the S-matrix and can only be generated by summing an infinite 
set of Feynman diagrams.   
Therefore, one is forced to face simultaneously
a plethora of technical difficulties, including
issues of regularization and renormalization,
nonrelativistic reductions, model dependence, and stability and
controllability of the approximation scheme.
Following the seminal work of Bethe and 
Salpeter~\cite{bet:51} many techniques and ideas have been developed, 
including numerical schemes, and have been applied to both QED and QCD with 
various degrees of success.  
More recently, the use of effective Lagrangians
has advanced the field considerably~\cite{recent_review},
and extensions to other fields of medium- and
high-energy physics are underway.  

In this paper we begin a program centered on the use of path
integrals for a systematic study of singular problems in quantum physics
and of the connection between field theory and
quantum mechanics {\it vis-\`{a}-vis\/} 
the description of bound states.
One of the main purposes of this approach is to formulate a
complementary point of view that may shed light on the 
difficult problems posed by bound states in quantum field theory.
It is to be expected that a more thorough 
understanding of these questions would eventually have a
substantial impact on the development of modern nuclear physics within 
the chiral Lagrangian approach~\cite{recent_review}, 
as well as on atomic and 
molecular physics~\cite{fra:71}.  
Even though the use of path integrals to study bound
states is hardly new, it has mainly been implemented within
the framework of quantum mechanics and, until recently, only for regular
potentials~\cite{kleinert,sav:99}. 
In this paper,
as we take the first steps in formulating this program,
we are confronted with the problem of describing singular potentials beyond
the Coulomb problem.  
In particular, we will concentrate upon the path-integral
treatment of the very important case of the inverse square potential,
$1/r^{2}$, the first-one with integer powers beyond the Coulomb case.  

The quantum-mechanical inverse square potential
has been extensively dealt with before via the
Schr\"{o}dinger equation~\cite{mot:49,gup:93,cam:00,cam:01}, 
and to a lesser extent
with path integrals~\cite{nel:64,pea:69,jar:88,bha:89,gro:98}. 
A recent proposal for a comprehensive treatment of this problem
was advanced within the Schr\"{o}dinger-equation approach, properly
combined with regularization and renormalization concepts 
borrowed from quantum field theory. This was 
done first for the one-dimensional case with a real-space regulator~\cite{gup:93};
later, it was generalized to a complete analysis of the $D$-dimensional case
using different regularization schemes~\cite{cam:00,cam:01}.
The current relevance of this problem stands out
not only from its field-theory treatment but also from its recent application
to the interaction between an electron and polar molecule, which was shown to be
a simple manifestation of a quantum anomaly~\cite{cam:01b}.

The main goals of the present paper are:

(i) To introduce a general framework for the
computation of bound states and other ``nonperturbative'' results
within the path-integral approach to quantum physics
(Sec.~\ref{sec:general_framework}).

(ii) To reformulate the problem of the inverse square potential
and reproduce the results of 
Ref.~\cite{cam:00} within the path-integral framework
(Secs.~\ref{sec:ISP}--\ref{sec:ISP_regularization}).

(iii) To generalize the inverse square potential
by the inclusion of delta-function interactions, with miscellaneous applications
(Sec.~\ref {sec:ISP+delta}).

(iv) To highlight the emergence of singular potentials within the effective-field theory
program 
and to sketch possible generalizations in quantum field theory
(Sec.~\ref{sec:QFT}).

The treatment of more singular potentials and the development of the 
full-fledged quantum-field-theory case is in progress and will be reported 
elsewhere.

\section{General Framework}
\label{sec:general_framework}

\subsection{Quantum-Mechanical Propagator}
\label{sec:QM_propagator}

Consider now a particle of mass $M$ 
subject to an interaction potential $V({\bf r},t)$ in  $D$-dimensions.
Its physics is completely described
by the quantum-mechanical propagator or transition amplitude
\begin{equation} 
 K_{D}  ( {\bf r}'', {\bf r}' ; t'',t' ) 
=
\left\langle 
{\bf r}''
\left| 
\hat{T}
\exp \left[
 -\frac{i}{\hbar} \int_{t'}^{t''}
\hat{H} dt 
\right]
\right|
{\bf r}'
 \right\rangle   
\; ,
\label{eq:propagator_QM_from_evolution_operator}
\end{equation}
where $\hat{T}$ is the time-ordering operator
and  $\hat{H}$ the Hamiltonian.
The propagator
admits the usual representation
\begin{equation}
K_{D}({\bf r}'', {\bf r}' ; t'',t') 
= 
\int_{  {\bf r} (t')  = {\bf r}'  }^{  {\bf r} (t'')  = {\bf r}'' }
 \;  
{\cal D} {\bf r} (t) \,
\exp \left\{ 
\frac{i}{\hbar} 
S \left[ {\bf r}(t)  \right]  ({\bf r}'', {\bf r}' ; t'',t')  
\right\}
\;  
\label{eq:propagator_QM}
\end{equation}
as a path integral, in which 
$S \left[ {\bf r}(t)  \right]  ({\bf r}'', {\bf r}' ; t'',t')  $ 
is the classical action functional
associated with ``paths''
$ {\bf r}(t)  $ that connect the end points
${\bf r} (t')  = {\bf r}',  {\bf r} (t'')  = {\bf r}'' $.
In anticipation of the derivations of Sec.~\ref{sec:infinite_summation},
we have carefully defined all the relevant variables in the functional dependence 
of 
$S \left[ {\bf r}(t)  \right]  ({\bf r}'', {\bf r}' ; t'',t')  $; specifically, 
the action is 
(i) a functional of 
$ {\bf r}(t)  $ and
(ii) a function of the end-point variables
$({\bf r}'', {\bf r}' ; t'',t')  $, 
for every chosen
$ {\bf r}(t)  $.

Equation~(\ref{eq:propagator_QM}) is
the primitive construct 
from which we will advance a general technique for the evaluation of the physics
of singular potentials. 
Furthermore,
we would like to explicitly show how to supplement this technique with regularization
{\it \`{a} la\/} field theory
for the derivation of the renormalized physics of the inverse square
potential---by itself and also combined with a delta-function interaction.

A careful analysis shows that one must exercise 
proper caution when evaluating Eq.~(\ref{eq:propagator_QM})
in non-Cartesian coordinates. For example,
when considering central potentials and 
transforming into hyperspherical coordinates in $D$ dimensions, 
one may need to consider extra terms of order $\hbar^{2}$ in the action.
These extra terms arise
whenever nonlinear transformations are implemented,
a fact that has been established in quantum mechanics~\cite{extra_terms_qm}
and quantum field theory~\cite{extra_terms_qft}.
However, these technical difficulties can be altogether circumvented by adopting a 
time-sliced expression in Cartesian coordinates
as the starting point in Eq.~(\ref{eq:propagator_QM}) and
transforming coordinates {\em before\/} taking the continuum limit.
Therefore, we introduce 
 a partition 
of the time $ T= t''-t' $ into $N$ 
equal 
discrete time units,
\begin{equation}
{\mathcal P}^{(N )} 
\left\{
t',t''
\right\}
:
\;
\left\{
t_{0}
 \equiv t',
t_{1},
\ldots
,
t_{N-1} ,
t_{N} \equiv t''
\right\}
\;  ,
\label{eq:time_partition}
\end{equation}
which yields a time lattice
of $(N+1)$ equidistant points
\begin{equation}
t_{j}= t'+j \epsilon   
\; \; \; \; \;        
 \epsilon = \frac{t''-t'}{N}
\; \; \; \; \;        
 (j=0,\ldots,N+1)        
\; ,
\end{equation}
with the end points fixed as in Eq.~(\ref{eq:time_partition}).
This procedure should be followed 
by taking the limit $N \rightarrow \infty$ only after all other calculations are completed.
Correspondingly, 
in Eq.~(\ref{eq:propagator_QM}), 
the action is replaced by its discrete counterpart
\begin{equation}
S^{(N)}  
 \left[ 
{\bf r}_{1},
\dots
{\bf r}_{N-1}
\right]  ({\bf r}'', {\bf r}' ; t'',t')  
=
\sum_{j=0}^{N-1}
\left[
\frac{  M  \left( {\bf r}_{j+1}
- {\bf r}_{j}  \right)^{2}}{2\epsilon} -
\epsilon V({\bf r}_{j}, t_{j})
\right]
\; ,
\label{eq:lattice_action}
\end{equation}
 in which
$ {\bf r}_{j} = {\bf r} (t_{j})$, while
the end points are
${\bf r}_{0} \equiv {\bf r}'$
and 
$ {\bf r}_{N} \equiv  {\bf  r}'' $.
It should be noticed that the variable $N$ has been isolated as a superscript 
in order to emphasize its distinct nature 
in the limit 
$
S^{(N)}  
 \left[ 
{\bf r}_{1},
\dots
{\bf r}_{N-1}
\right]  ({\bf r}'', {\bf r}' ; t'',t')  
\rightarrow
S \left[ {\bf r}(t)  \right]  ({\bf r}'', {\bf r}' ; t'',t')  
$,
with the variables 
$\left\{ {\bf r}_{1},
\dots
{\bf r}_{N-1}
\right\}$ 
asymptotically generating a ``path'' ${\bf r}(t)$.
Then,  Eq.~(\ref{eq:propagator_QM}) can be interpreted as a shorthand for the
time-sliced expression
\begin{equation}
K_{D}({\bf r}'', {\bf r}' ; t'',t') 
= 
\lim_{N \rightarrow \infty}
\int_{  {\bf r} (t')  = {\bf r}'  }^{  {\bf r} (t'')  = {\bf r}'' }
 \;  
{\cal D} ^{(N)} {\bf r} (t)
\,
\exp \left\{
\frac{i}{\hbar} 
S^{(N)}  
 \left[ 
{\bf r}_{1},
\dots
{\bf r}_{N-1}
\right]  ({\bf r}'', {\bf r}' ; t'',t')  
\right\}
\;  ,
\label{eq:propagator_QM2}
\end{equation}
where the integration and measure symbols stand for 
the formal $(N-1)$-fold integral operation
\begin{equation}
\int_{  {\bf r} (t')  = {\bf r}'  }^{  {\bf r} (t'')  = {\bf r}'' }
 \;  
{\cal D} ^{(N)} {\bf r} (t)
\equiv
\left[
{\mathcal C} (M,\epsilon,\hbar,D) 
\right]^{N}
\,
\left[
\prod_{k=1}^{N-1} \int_{ \mathbb{R}^{D} } d^{D} {\bf r}_{k} 
\right]
\;  ,
\label{eq:measure_and_integral}
\end{equation}
with
\begin{equation}
{\mathcal C} (M,\epsilon,\hbar,D) 
=
\left(
\frac{M}{2 \pi i \epsilon \hbar} 
\right)^{D/2}
\;  .
\label{eq:Feynman_measure_factor}
\end{equation}

As is well known, 
the propagator $K_{D}({\bf r}'', {\bf r}' ; t'',t') $
solves the initial-value problem for the 
Schr\"{o}dinger equation, permitting the computation of the wave function
$\Psi ({\bf r},t)$ at any time $t$, given 
a particular-time value $\Psi ({\bf r},t_{0})$.
A related quantity is the retarded (causal) Green's function
$G_{D}({\bf r}'', {\bf r}' ; t'',t') $, which solves the 
Schr\"{o}dinger equation with an additional arbitrary 
inhomogeneous term, and which is 
related to the propagator by
 \begin{equation}
G_{D}({\bf r}'', {\bf r}' ; t'',t') 
=
\theta (t''-t')
\,
K_{D}({\bf r}'', {\bf r}' ; t'',t') 
\;   ,
\label{eq:causal_Green}
\end{equation}
where
$\theta (T)$ stands for the Heaviside function.

In this paper we
 will consider time-independent singular potentials,
with special emphasis on the inverse square potential and the delta-function
interaction.
These particular cases
are included in the class of time-independent
Hamiltonians, for which the lattice action
\begin{equation}
S^{(N)}  
 \left[ 
{\bf r}_{1},
\dots
{\bf r}_{N-1}
\right]  ({\bf r}'', {\bf r}' ; t'',t')  
 \equiv 
S^{(N)}  
 \left[ 
{\bf r}_{1},
\dots
{\bf r}_{N-1}
\right]  ({\bf r}'', {\bf r}' ; t''-t')  
\; ,
\end{equation}
as well as its continuum limit
\begin{equation}
S \left[ {\bf r}(t)  \right]  ({\bf r}'', {\bf r}' ; t'',t')  
 \equiv 
S \left[ {\bf r}(t)  \right]  ({\bf r}'', {\bf r}' ; t''-t')  
\;  ,
\end{equation}
and the propagator
\begin{equation}
 K_{D}  ( {\bf r}'', {\bf r}' ; t'',t' )  
 \equiv  
K_{D}  ( {\bf r}'', {\bf r}' ; t''-t')  
\;
\end{equation}
are functions of the end-point times
$t'$ and $t''$
only through the difference $T=t''-t'$.
Under this assumption, it
is also possible to specialize to ``paths'' 
connecting the end points
${\bf r} (0)  = {\bf r}',  {\bf r} (T)  = {\bf r}'' $.
Then, the Fourier transform of $G_{D} ({\bf r}'', {\bf r}' ; T) $
with respect to $T$ defines
the corresponding energy Green's function
\begin{eqnarray}
G_{D} ({\bf r}'', {\bf r}' ; E) 
& = &
\frac{1}{ i\hbar }
\int_{-\infty}^{\infty} dT  
e^{iET/\hbar}
G_{D}({\bf r}'', {\bf r}' ; T)  
\nonumber \\
& = &
\frac{1}{ i\hbar }
\int_{0}^{\infty} dT  
e^{iET/\hbar}
K_{D}({\bf r}'', {\bf r}' ; T)  
\label{eq:GreenF_def}
 \\
& = &
\frac{1}{ i\hbar }
\int_{0}^{\infty} dT  
\,
\int_{  {\bf r} (0)  = {\bf r}'  }^{  {\bf r} (T)  = {\bf r}'' }
 \;  
{\cal D} {\bf r} (t) \,
\exp 
\left(
\frac{i}{\hbar} 
\left\{ 
S \left[ {\bf r}(t)  \right]  ({\bf r}'', {\bf r}' ; T)  +
ET 
\right\}
\right)
\; .
\label{eq:GreenF_def_from_PIs}
\end{eqnarray}
The energy Green's function
is a convenient tool for spectral analyses
because it satisfies the
operator relation
\begin{equation}
G_{D} ({\bf r}'', {\bf r}' ; E) 
=
\left\langle 
{\bf r}'' 
\left|
\left( 
E - \hat{H} + i \epsilon 
\right)^{-1}
 \right|
{\bf r}'
 \right\rangle   
\; ,
\label{eq:GreenF_operator_def}
\end{equation}
where $i\epsilon$ is a small positive imaginary part---this follows by combining
Eqs.~(\ref{eq:propagator_QM_from_evolution_operator})
 and (\ref{eq:GreenF_def}).
In particular, it
admits a spectral representation,
\begin{equation}
G_{D} ({\bf r}'', {\bf r}' ; E) 
=
\sum_{n}
\frac{ 
\psi_{n} ({\bf r}'')  \, \psi_{n} ^{*}({\bf r}')  
}{E-E_{n} +i \epsilon}
+
\int d \alpha
\,
\frac{
\psi_{\alpha} ({\bf r}'')  \, \psi_{\alpha} ^{*}({\bf r}')  
}{E-E_{\alpha} +i \epsilon}
\;  ,
\label{eq:energy_propagator_states}
\end{equation}
which identifies
the bound states as poles and the scattering states 
as branch cuts, as well as the corresponding
stationary wave functions~\cite{kleinert}.
For example,
for a nondegenerate bound state of energy $E_{n}$,
the wave function
$\psi_{n}({\bf r}) =
|\psi_{n}({\bf r}) |
e^{i \delta_{n}({\bf r})}$
can be uniquely extracted, up to an arbitrary global phase factor, from
\begin{equation}
|\psi_{n}({\bf r}) |
=
\left|
{\rm Res} \left\{ G({\bf r},{\bf r}; E_{n} 
\right\}
\right|^{1/2}
\;  
\label{eq:residue_magnitude_wf}
\end{equation}
and
\begin{equation}
\delta_{n}({\bf r}) 
=
- i \ln
\left\{
\frac{
{\rm Res} \left\{ G({\bf r},{\bf r}_{0}; E_{n} ) \right\}
}{
\left[
{\rm Res} \left\{ G({\bf r},{\bf r}; E_{n} )\right\}
\,
{\rm Res} \left\{ G({\bf r}_{0},{\bf r}_{0}; E_{n} )\right\}
\right]^{1/2}
}
\right\}
\;  ,
\label{eq:residue_phase_wf}
\end{equation}
where ${\bf r}_{0}$ is a convenient reference
point and ${\rm Res} \{ f(z_{0}) \}$ represents the residue of the function
$f(z)$ at the pole $z=z_{0}$.

In what follows 
our arguments and derivations will be based on Eqs.~(\ref{eq:propagator_QM2})
and (\ref{eq:GreenF_def_from_PIs}),
as well as on their counterparts in hyperspherical coordinates, which we will 
consider next.

\subsection{Central Interactions}
\label{sec:central_interactions}

For the all-important class of central potentials,
Eq.~(\ref{eq:propagator_QM2}) can be conveniently rewritten 
in hyperspherical  polar coordinates~\cite{cam:00,cam:01,erd:53}.
This can be accomplished by separation of variables, 
a procedure that can be systematically
implemented by repeated use of the addition formula~\cite{wat:44}
$e^{iz\cos \psi}
=
(iz/2)^{-\nu} \Gamma (\nu)
\,
 \sum_{l=0}^{\infty} 
(l+\nu)
I_{l+\nu} (iz ) C_{l}^{(\nu)} (\cos \psi )$,
where $I_{p}(x)$ 
is the modified Bessel function of the first kind and order $p$,
$C_{l}^{(\nu)}(x)$ is a Gegenbauer polynomial, and
$\nu =D/2 -1$ is a related dimensionality parameter.
In turn, the Gegenbauer polynomials can be resolved into 
hyperspherical harmonics
$Y_{l m} ({\bf \Omega})$, 
labeled with the $D$-dimensional angular momentum quantum numbers
$l$ and $m$, where the latter has a multiplicity
 $d_{l}= (2l+D-2)(l+D-3)!/l!(D-2)!$~\cite{erd:53}.
As a result,
the propagator admits the
partial-wave expansions
\begin{eqnarray}
K_{D}({\bf r}'', {\bf r}' ; T) 
& = &
\frac{\Gamma (\nu)}{2 \pi^{D/2} }
\,
\left( r'' r' \right)^{-(D-1)/2}
\,
\sum_{l= 0}^{\infty}
\,
(l+\nu)
\,
C_{l}^{(\nu)} (\cos \psi_{ {\bf \Omega}'', {\bf \Omega}' } ) 
\,
K_{l +\nu}(r'',r';T)
\nonumber
\\
& = &
\left( r'' r' \right)^{-(D-1)/2}
\,
\sum_{l= 0}^{\infty}
\sum_{m=1}^{d_{l}}
Y_{l m} ({\bf \Omega}'')
Y_{l m}^{ *} ({\bf \Omega}')
\,
K_{l +\nu}(r'',r';T)
\;  ,
\label{eq:propagator_partial_wave_exp}
\end{eqnarray}
where $\cos \psi_{ \, {\bf \Omega}'', {\bf \Omega}' } = {\bf \hat{r}}'' \cdot {\bf \hat{r}}' $ and
the radial prefactors are conveniently written so that,
for each angular momentum channel $l$,  
the radial propagator
$K_{l+\nu}(r'',r';T)$ satisfies the composition property
\begin{equation}
K_{l+\nu}(r'',r';t''-t')
=
\int_{0}^{\infty}
dr
\,
K_{l+\nu}(r'',r;t''-t)
K_{l+\nu}(r,r';t-t')
\;  .
\end{equation}
Moreover,
the radial propagator in
Eq.~(\ref{eq:propagator_partial_wave_exp}) 
can be explicitly evaluated by means of the
path-integral expression
\begin{eqnarray}
K_{l+\nu}(r'',r';T)  = 
\lim_{N \rightarrow \infty}
\left( \frac{M}{2 \pi i \epsilon \hbar} \right)^{N/2}
\,
&   &
\prod_{k=1}^{N-1}
\left[ 
 \int_{0}^{\infty} d r_{k}
\right]
\;  
\mu_{l+\nu}^{(N)} [r^{2}] 
\,
\nonumber  \\
&  \times  &
\exp \left\{
\frac{i}{\hbar} 
R^{(N)}  
 \left[ 
 r_{1},
\dots
 r_{N-1}
\right]  (r'', r' ; T)  
\right\}
\;   ,
\label{eq:propagator_QM_spherical_coords_2D_expansion}
\end{eqnarray}
where the radial action is
\begin{equation}
R^{(N)}  
 \left[ 
 r_{1},
\dots
 r_{N-1}
\right]  (r'', r' ; T)  
=
\sum_{j=0}^{N-1}
\left[
\frac{  M
\left( r_{j+1}
- r_{j}  \right)^{2}  
}{2\epsilon} 
-
\epsilon V( r_{j} )
\right]
\; .
\end{equation}
In Eq.~(\ref{eq:propagator_QM_spherical_coords_2D_expansion})
a radial functional weight 
\begin{equation}
\mu_{l+\nu}^{(N)} [r^{2}]
=
\prod_{j=0}^{N-1} 
\left[ 
\sqrt{2 \pi z_{j}} e^{-z_{j}}
I_{l+\nu} (z_{j})
\right]
\; 
\label{eq:radial_pseudomeasure} 
\end{equation}
has been properly
defined with the radial variables appearing through the characteristic 
dimensionless ratio
\begin{equation}
z_{j}
=
 \frac{M r_{j} r_{j+1}}{ i \epsilon \hbar}
\;  .
\label{eq:z_variable}
\end{equation}
As a consequence, 
the radial path integral, supplemented by the condition $r(t) \geq 0$,
can be given a formal continuum representation
\begin{equation}
K_{l+\nu}(r'',r';T)
=
\int_{ r(0) = r'}^{ r(T)=r'' }
{\cal D}  r(t)  \,
\; 
\mu_{l+ \nu} [r^{2}]
\exp \left(
\frac{i}{\hbar} 
\int_{0}^{T}
dt
\left\{
\frac{M}{2} \, \left[   \stackrel{.}{r}  (t) \right]^{2}
-
V(r(t))
\right\}
\right)
\;   
\label{eq:propagator_QM_spherical_coords_2D_expansion_continuous}
\end{equation}
in terms of the usual one-dimensional path-integral pseudomeasure 
${\cal D}  r(t)  $.
However, our subsequent analysis will be based on the time-sliced
Eq.~(\ref{eq:propagator_QM_spherical_coords_2D_expansion}),
which has been called a Besselian path integral~\cite{gro:98} because of the appearance of
modified Bessel functions in the pseudomeasure~(\ref{eq:radial_pseudomeasure}).
A similar approach can be applied to the energy Green's function
$G_{D} ({\bf r}'', {\bf r}' ; E) $,
which through a partial-wave expansion of the form~(\ref{eq:propagator_partial_wave_exp})
defines a radial counterpart
\begin{equation}
G_{l +\nu}(r'',r';E)
 = 
\frac{1}{ i\hbar }
\int_{0}^{\infty} dT  
e^{iET/\hbar}
K_{l +\nu}(r'',r';T)
\; .
\label{eq:radial_GreenF_def}
\end{equation}
It should be noticed that
Eqs.~(\ref{eq:propagator_QM_spherical_coords_2D_expansion})--(\ref{eq:radial_GreenF_def})
explicitly exhibit  a property that may be called interdimensional 
dependence~\cite{interdimensional}:
$D$ and $l$ appear in the combination $l+\nu$.

It is well known that the path-integral 
expression~(\ref{eq:propagator_QM_spherical_coords_2D_expansion})
for the free-particle radial propagator
can be computed analytically and exactly
(even before taking the continuum limit, $N \rightarrow \infty$),
with the familiar result~\cite{pea:69}
\begin{eqnarray}
K^{(0)}_{l+\nu} (r'',r';T) 
=
\frac{M}{i \hbar T} 
\,
\sqrt{ r' r''} 
\,
\exp \left[ \frac{i M}{2 \hbar T} \left(r^{\prime 2} + r^{\prime \prime 2} 
\right) \right]
I_{l+\nu} \left( \frac{Mr'r''}{i \hbar T} \right)
\;   ;
\label{eq:propagator_free_particle}
\end{eqnarray}
this is accomplished through the use of 
Bessel-function identities and recursion relations~\cite{weber}. 
In addition, the Fourier transform~(\ref{eq:radial_GreenF_def}) of 
$
K^{(0)}_{l+\nu} (r'',r';T) 
$~\cite{gra:00}
provides a closed analytical expression
for the free-particle radial energy Green's function
\begin{equation}
G^{(0)}_{l+\nu} (r'',r';E)    
 = 
 -
\frac{2M}{\hbar^{2}} 
\,
\sqrt{ r'r''} 
\,
I_{l+\nu}( \kappa r_{<} ) K_{l+ \nu} (\kappa r_{>} )
\; ,
\label{eq:GreenF_free_particle}
\end{equation}
in which the energy variable appears in the 
combination
\begin{equation}
\kappa=
  \frac{ \sqrt{-2ME} }{\hbar}
\;  .
\label{eq:imaginary_wave_number}
\end{equation}
As usual, in Eq.~(\ref{eq:GreenF_free_particle}),
$r_{<}$ ($r_{>}$) is the smaller 
(larger) of $r'$ and $r''$, while 
$I_{p}(x)$ 
and
$K_{p}(x)$ 
are the modified Bessel functions of the first kind and second kind
(and order $p$) respectively.

Explicit evaluation of 
Eq.~(\ref{eq:propagator_QM_spherical_coords_2D_expansion}) 
beyond the free-particle case has been accomplished only in a few 
special cases, including the radial harmonic oscillator~\cite{gro:98},
using miscellaneous limiting procedures in the limit
$N \rightarrow \infty$.
For the case of more general potentials, 
we will next describe a 
 technique that permits the summation of perturbation-theory
contributions to all orders.

\section{Infinite Summation of Perturbation Theory}
\label{sec:infinite_summation}

In this section we introduce a general 
technique that consists of a perturbation 
expansion followed by its summation to all orders---this is
essentially the ``perturbation approach'' pioneered in Refs.~\cite{bha:89,bha:88}.
The objective of this combined procedure is to extract ``nonperturbative''
results, including the emergence of bound states (whenever appropriate), from
a manifestly perturbative scheme.
This is only possible by summing the perturbative series to all orders.
We generalize the results of Refs.~\cite{bha:89,bha:88}  in a number of ways:
(i) we present the theory in any number of dimensions,
 for the full-fledged propagator and for the Green's function;
(ii) we develop it within the time-sliced formulation of path integrals,
with emphasis upon possible subtleties involved in nonlinear
coordinate changes;
and (iii) we explicitly show that the expansion for radial path integrals 
does not require any additional modifications, provided that the Besselian
pseudomeasure~(\ref{eq:radial_pseudomeasure}) be used at every order.
At the very least,
this technique will play a crucial role
in providing a complete and satisfactory derivation of the
renormalized physics of the inverse square potential and delta-function interactions.

\subsection{Perturbation Theory}
\label{sec:perturbation_theory}

This procedure is based on the standard 
resolution of the action in the form
\begin{eqnarray}
S 
\left[ {\bf r}(t)  \right]  ({\bf r}'', {\bf r}' ; t'',t')  
& = &
S^{(0)}   
\left[ {\bf r}(t)  \right]  ({\bf r}'', {\bf r}' ; t'',t')  
+
\tilde{S}  
\left[ {\bf r}(t)  \right]  ({\bf r}'', {\bf r}' ; t'',t')  
\nonumber \\
& = &
S^{(0)}   
\left[ {\bf r}(t)  \right]  ({\bf r}'', {\bf r}' ; t'',t')  
-
\int_{t'}^{t''} 
dt \;
V({\bf r} (t),t)
\;  ,
\label{eq:Hamiltonian_perturb_split}
\end{eqnarray}
where
$
S^{(0)}  \left[ {\bf r}(t)  \right]  ({\bf r}'', {\bf r}' ; t'',t')  
$ 
is the action for a  problem whose
propagator is already known
(for example, the free-particle action) while
$
\tilde{S}  \left[ {\bf r}(t)  \right]  ({\bf r}'', {\bf r}' ; t'',t')  
$ 
provides the additional interactions 
$
V({\bf r},t )
$ 
in 
$
S  \left[ {\bf r}(t)  \right]  ({\bf r}'', {\bf r}' ; t'',t')  
$. 

Even though this beginning step~(\ref{eq:Hamiltonian_perturb_split})
looks like standard perturbation theory, it is
in fact intended as an exact rearrangement based on infinite summations 
performed to all orders. 
This resolution can be applied to any path-integral expression:
to Eq.~(\ref{eq:propagator_QM2}), for the full-fledged $D$-dimensional problem;
 and to Eq.~(\ref{eq:propagator_QM_spherical_coords_2D_expansion}),
for the radial propagator.
Just as before,  the time-lattice versions guarantee that all subtleties are 
properly accounted for in this formulation. Thus, we will start 
with a lattice action described by the
generic Eqs.~(\ref{eq:lattice_action})--(\ref{eq:Feynman_measure_factor}),
as defined by the associated partition~(\ref{eq:time_partition}),
i.e.,
\begin{eqnarray}
S^{(N)}  
 \left[ 
{\bf r}_{1},
\dots
{\bf r}_{N-1}
\right]  ({\bf r}'', {\bf r}' ; t'',t')  
& = &
S^{(0;N)}   
 \left[ 
{\bf r}_{1},
\dots
{\bf r}_{N-1}
\right]  ({\bf r}'', {\bf r}' ; t'',t')  
+
\tilde{S}^{(N)}  
 \left[ 
{\bf r}_{1},
\dots
{\bf r}_{N-1}
\right]  ({\bf r}'', {\bf r}' ; t'',t')  
\nonumber \\
& = &
S^{(0;N)}   
 \left[ 
{\bf r}_{1},
\dots
{\bf r}_{N-1}
\right]  ({\bf r}'', {\bf r}' ; t'',t')  
- \epsilon
\sum_{j = 0}^{N-1} 
V({\bf r}_{j},t_{j})
\;  
\label{eq:Hamiltonian_perturb_split_lattice}
\end{eqnarray}
in lieu of Eq.~(\ref{eq:Hamiltonian_perturb_split}),
where $S^{(0;N)}   $ 
stands for the corresponding lattice unperturbed action.
In each one of these path integrals,
the exponential $e^{i\tilde{S}^{(N)}/\hbar}$ is to be expanded
to all orders,
\begin{eqnarray}
&  & 
\exp
\left\{
\frac{i}{\hbar}
\tilde{S}^{(N)}  
 \left[ 
{\bf r}_{1},
\dots
{\bf r}_{N-1}
\right]  ({\bf r}'', {\bf r}' ; t'',t') 
\right\}
 \nonumber 
\\
& = &
\sum_{n=0}^{\infty}
\frac{\epsilon^{n}}{n!}
\;
\left[
\sum_{j_{n}=0}^{N-1}
\,
\ldots
\,
\sum_{j_{1}=0}^{N-1}
\,
\frac{ V( {\bf r}_{j_{n}}, t_{j_{n}} ) }{i \hbar}
\ldots
\frac{ V( {\bf r}_{j_{1}}, t_{j_{1}} ) }{i \hbar}
\right]
\nonumber
\\
 & = &
\sum_{n=0}^{\infty}
\epsilon^{n}
\;
\left[
\sum^{N-1}_{ 
          j_{1} < j_{2} <  \dots  < j_{n} =0 }
\,
\frac{ V( {\bf r}_{j_{n}}, t_{j_{n}} ) }{i \hbar}
\ldots
\frac{ V( {\bf r}_{j_{1}}, t_{j_{1}} ) }{i \hbar}
\right]
\,
\left[ 1 + O \left( \frac{1}{N} \right) \right]
\;  ,
\label{eq:perturbation_potential_expansion}
\end{eqnarray}
where corrections of order 
$ O \left( 1/N \right) $ are associated with the terms
with repeated indices.
In order to simplify the notation, while at the same time preventing ambiguities,
let us define
\begin{eqnarray}
t_{(0)}
& \equiv &
t_{0}
=
t'
\; \; \; \; \; \; \; 
t_{(n+1)}
\equiv 
t_{N}
=
t''
\nonumber \\
t_{(\alpha)}
& \equiv &
t_{j_{\alpha}}
\; \; \; \; \; \; \; 
{\bf r}_{(\alpha)}
\equiv 
{\bf r}_{j_{\alpha}}
\; \; \; \; \; \; \; 
(\alpha = 1, \dots , n)
\;  .
\end{eqnarray}
Then, {\em at each order of perturbation theory\/},
Eq.~(\ref{eq:perturbation_potential_expansion}) defines a time ordering
with which we can associate the partition
\begin{equation}
{\mathcal P}^{(n+1)} 
\left\{
t',t''
\right\}
:
\left\{
t_{(0)} \equiv t', 
t_{(1)},
t_{(2)},
\ldots
,
t_{(n)},
t_{(n+1)} \equiv t''
\right\}
\;  
\label{eq:time_interval_partition}
\end{equation}
leading to a resolution of the unperturbed action
\begin{eqnarray}
S^{(0;N)}   
 \left[ 
{\bf r}_{1},
\dots
{\bf r}_{N-1}
\right]  ({\bf r}'', {\bf r}' ; t'',t')  
& \equiv &
S^{(0;N)}   
 \left[ 
{\bf r}_{1},
\dots
{\bf r}_{N-1}
\right]  ({\bf r}'', {\bf r}' ; t'',t')  
\nonumber \\
& = &
\sum_{\alpha=0}^{n}
S^{(0;N)}   
 \left[ 
{\bf r}_{1},
\dots
{\bf r}_{N-1}
\right]  ({\bf r}_{(\alpha+1)}, {\bf r}_{(\alpha)} ; t_{(\alpha+1)},t_{(\alpha)} )  
\;  .
\label{eq:perturbative_resolution_of_unperturbed_action}
\end{eqnarray}
In Eq.~(\ref{eq:perturbative_resolution_of_unperturbed_action})
the whole interval $T=t''-t'$ has been divided into
$n+1$ subintervals
 $T_{(\alpha)}=
t_{(\alpha+1)}- t_{(\alpha)}$,
by introducing the points $t_{(\alpha)}$ with $\alpha=0, \ldots, n+1$,
at each order $n$ in perturbation theory.
Correspondingly,
the $D$-dimensional propagator becomes
\begin{equation}
K_{D}({\bf r}'', {\bf r}' ; t'',t')
= 
\sum_{n=0}^{\infty}
K^{(n)}_{D}({\bf r}'', {\bf r}' ; t'',t')
\; ,
\label{eq:propagator_perturb_expansion}
\end{equation}
in which
the $n$-th order contribution is
\begin{eqnarray}
K^{(n)}_{D}({\bf r}'', {\bf r}' ; t'',t')
&
=
& 
\lim_{N \rightarrow \infty}
\,
\int_{  {\bf r} (t')  = {\bf r}'  }^{  {\bf r} (t'')  = {\bf r}'' }
 \;  
{\cal D} ^{(N)} {\bf r} (t) 
\,
\nonumber \\
& \times &
\epsilon^{n}
\;
\left.
\left[
\sum^{N-1}_{ 
          j_{1} < j_{2} <  \dots  < j_{n} =0 }
\,
\frac{ V( {\bf r}_{j_{n}}, t_{j_{n}} ) }{i \hbar}
\ldots
\frac{ V( {\bf r}_{j_{1}}, t_{j_{1}} ) }{i \hbar}
\right]
\,
\left[ 1 + O \left( \frac{1}{N} \right) \right]
\right|_{ \left\{
t_{j_{\alpha}}
 \equiv 
t_{(\alpha)}
;
{\bf r}_{j_{\alpha}} 
 \equiv 
{\bf r}_{(\alpha)}
\right\}
}
\nonumber \\
& \times  &
\prod_{\beta=0}^{n}
\left[
 \exp \left\{ \frac{i}{\hbar} 
S^{(0;N)}   
 \left[ 
{\bf r}_{1},
\dots
{\bf r}_{N-1}
\right]  
\left( {\bf r}_{(\beta+1)}, {\bf r}_{(\beta)} ; t_{(\beta+1)},t_{(\beta)}  \right)  
 \right\}
\right]
\; ,
\label{eq:straight_propagator_perturb_expansion_nth_order}
\end{eqnarray}
with the time-sliced measure
displayed in Eqs.~(\ref{eq:measure_and_integral})
and (\ref{eq:Feynman_measure_factor}).
Finally, before taking the limit $N \rightarrow \infty$
in Eq.~(\ref{eq:straight_propagator_perturb_expansion_nth_order}),
it is suitable to assume that $N \gg n$, followed by a rearrangement
of the time lattice according to the following
scheme:

(i) 
The whole time interval $T=t''-t'$ is divided into $n+1$ subintervals
just as in Eq.~(\ref{eq:perturbative_resolution_of_unperturbed_action}).

(ii)
Each subinterval $[t_{(\alpha)},t_{(\alpha+1)}]$ is
partitioned in the usual way into $N_{\alpha}$ parts,
just as in Eq.~(\ref{eq:time_partition}), i.e.,
\begin{equation}
{\mathcal P}^{(N_{ \alpha}) } 
\left\{
t_{(\alpha)},t_{(\alpha+1)}
\right\}
:
\left\{
t_{\alpha,j_{0}}
 \equiv t_{(\alpha)},
t_{\alpha,j_{1}},
\ldots
,
t_{\alpha,j_{N_{\alpha}-1} },
t_{\alpha,j_{N_{\alpha}} }
 \equiv t_{(\alpha+1)}
\right\}
\;  .
\label{eq:time_subinterval_partition}
\end{equation}

(iii) Therefore, the whole interval consists of the following sub-partitions
\begin{equation}
{\mathcal P}^{(N) } 
\left\{
t',t''
\right\}
=
\left\{
{\mathcal P}^{(N_{ 0}) } 
\left\{
t', t_{1}
\right\};
{\mathcal P}^{(N_{ 1}) } 
\left\{
t_{1}, t_{2}
\right\};
\ldots
;
{\mathcal P}^{(N_{ n}) } 
\left\{
t_{n}, t''
\right\}
\right\}
\;  ,
\label{eq:time_global_partition}
\end{equation}
where the notational redefinition implied in Eq.~(\ref{eq:time_subinterval_partition})
makes the parentheses surrounding the subscript $\alpha$ 
redundant. Thus, 
 the omission of these parentheses amounts to the replacements
\begin{eqnarray}
t_{(\alpha) } 
& \hookrightarrow &
t_{\alpha }
\\
{\bf r}_{(\alpha)}
& \hookrightarrow &
{\bf r}_{\alpha}
\;  ,
\end{eqnarray}
which will be assumed in our subsequent discussion.
With the partition defined by  Eqs.~(\ref{eq:time_subinterval_partition})
and (\ref{eq:time_global_partition}),
a partial measure
${\cal D} ^{(N_{\alpha})} {\bf r} (t) $ is 
defined just as in Eq.~(\ref{eq:measure_and_integral}), with the result
\begin{equation}
\int_{  {\bf r} (t')  = {\bf r}'  }^{  {\bf r} (t'')  = {\bf r}'' }
\,
{\cal D} ^{(N)} {\bf r} (t) 
=
\prod_{\alpha=1}^{n}
 \left[ \int
 d^{D} 
{\bf r}_{\alpha}
\right]
\;
\prod_{\beta=0}^{n} 
\left[
\int_{  {\bf r} (t_{\beta})  = {\bf r}_{\beta}  }^{  {\bf r} (t_{\beta+1})  = {\bf r}_{\beta+1} }
{\cal D} ^{(N_{\beta})} {\bf r} (t) 
\right]
\;  ,
\label{eq:partition_of_measure_and_integral}
\end{equation}
where
the first set of integration factors represents the contribution from 
the end points
of each subinterval---conveniently singled out  
in order to avoid overcounting.
Remarkably,
in Eq.~(\ref{eq:partition_of_measure_and_integral}),
the factors ${\mathcal C} (M,\epsilon,\hbar,D) $
[from Eqs.~(\ref{eq:measure_and_integral}) and
(\ref{eq:Feynman_measure_factor})]
naturally redistribute in such a way that each subinterval inherits the proper measure.

Finally, in the limit $N \rightarrow \infty$,
\begin{eqnarray}
& &
\epsilon^{n}
\;
\left[
\sum^{N-1}_{ 
          j_{1} < j_{2} <  \dots  < j_{n} =0 }
\,
\frac{ V( {\bf r}_{j_{n}}, t_{j_{n}} ) }{i \hbar}
\ldots
\frac{ V( {\bf r}_{j_{1}}, t_{j_{1}} ) }{i \hbar}
\right]
\,
\left[ 1 + O \left( \frac{1}{N} \right) \right]
\;  
\nonumber \\
&  &
\stackrel{ \; (N \rightarrow \infty) \; }{\longrightarrow}
\,
\int_{t'}^{t''} 
dt_{n} 
\int_{t'}^{t_{n}} 
dt_{n-1} 
\,
\ldots
\,
\int_{t'}^{t_{3}} 
dt_{2} 
\int_{t'}^{t_{2}} 
dt_{1} 
\;
\frac{ V({\bf r}_{n}, t_{n})
                 }{i \hbar}
\ldots
\frac{ V({\bf r}_{1}, t_{1})
                } {i \hbar}
\;  .
\label{eq:perturbation_potential_expansion_T_ordering}
\end{eqnarray}
Then,
as a result of the integration measure~(\ref{eq:partition_of_measure_and_integral}),
Eq.~(\ref{eq:straight_propagator_perturb_expansion_nth_order}) takes the form
\begin{eqnarray}
K^{(n)}_{D}({\bf r}'', {\bf r}' ; t'',t')
& = &
\int_{t'}^{t''} dt_{n} 
\int_{t'}^{t_{n}} dt_{n-1} 
\ldots
\int_{t'}^{t_{2}} dt_{1} 
\nonumber 
\\
& \times &
\prod_{\alpha=1}^{n} \left[
 \int
d^{D} {\bf r}_{\alpha} \,
\frac{   V({\bf r}_{\alpha},t_{\alpha})  
                       }{i \hbar} 
\right]
\,
\prod_{\beta=0}^{n} 
\left[
K^{(0)}_{D}({\bf r}_{\beta + 1}, {\bf r}_{\beta} ; 
t_{\beta+1},t_{\beta} )  \right]
\;   ,
\label{eq:propagator_perturb_expansion_nth_order_general}
\end{eqnarray}
where the contributions from 
$
S^{(0)}   
 \left[ 
{\bf r} (t)
\right]  
\left( {\bf r}_{\beta+1}, {\bf r}_{\beta} ; t_{\beta+1},t_{\beta}  \right)  
$
have been converted into the propagators
$K^{(0)}_{D}({\bf r}_{\beta + 1}, {\bf r}_{\beta} ; T_{\beta} )  $,
as $N \rightarrow \infty$.
For the particular case of time-independent Hamiltonians, 
Eq.~(\ref{eq:propagator_perturb_expansion_nth_order_general})
can be recast into the
alternative expressions
\begin{eqnarray}
K^{(n)}_{D}({\bf r}'', {\bf r}' ; T)
& = &
\prod_{\alpha=1}^{n} \left[
 \int
d^{D} {\bf r}_{\alpha} \,
\frac{   V({\bf r}_{\alpha})  }{i \hbar} 
\right]
\nonumber 
\\
& \times &
\int_{t'}^{t''} dt_{n} 
\int_{t'}^{t_{n}} dt_{n-1} 
\ldots
\int_{t'}^{t_{2}} dt_{1} 
\,
\prod_{\beta=0}^{n} 
\left[
K^{(0)}_{D}({\bf r}_{\beta + 1}, {\bf r}_{\beta} ; 
t_{\beta+1}-t_{\beta} )  \right]
\;   
\label{eq:propagator_perturb_expansion_nth_order}
\end{eqnarray}
and
\begin{equation}
K^{(n)}_{D}({\bf r}'', {\bf r}' ; T)
=
\prod_{\alpha=1}^{n} \left[
 \int
d^{D} {\bf r}_{\alpha} \,
\frac{   V({\bf r}_{\alpha})  }{i \hbar} 
\right]
\,
\prod_{\beta=0}^{n} \left[ \int_{0}^{\infty} 
d T_{\beta}  \,  
K^{(0)}_{D}({\bf r}_{\beta + 1}, {\bf r}_{\beta} ; T_{\beta} )  \right]
\,
\!
\delta
 \left(  
T - 
\sum_{\gamma=0}^{n} T_{\gamma}
 \right)
\;   ;
\label{eq:propagator_perturb_expansion_nth_order_2}
\end{equation}
correspondingly, from Eq.~(\ref{eq:propagator_perturb_expansion_nth_order_2}),
the Fourier transform
defining
the energy Green's function yields the result
\begin{equation}
G_{D}({\bf r}'', {\bf r}' ; E) 
=
\sum_{n=0}^{\infty}
G^{(n)}_{D}({\bf r}'', {\bf r}' ; E) 
\;  ,
\label{eq:GreenF_perturb_expansion}
\end{equation}
with an $n$-th order contribution
\begin{equation}
G^{(n)}_{D}({\bf r}'', {\bf r}' ; E) 
=
\prod_{\alpha=1}^{n} 
\left[ 
\int_{0}^{\infty}
d^{D} {\bf r}_{\alpha} \,
 V({\bf r}_{\alpha})  
\right]
\,
\prod_{\beta=0}^{n} \left[ 
G^{(0)}_{D}({\bf r}_{\beta + 1}, {\bf r}_{\beta} ; E )  \right]
\;  .
\label{eq:GreenF_perturb_expansion_nth_order}
\end{equation}
Direct comparison of 
Eqs.~(\ref{eq:propagator_perturb_expansion_nth_order}) 
and (\ref{eq:GreenF_perturb_expansion_nth_order})
shows that the former displays characteristic extra integrations with respect to time;
these arise when separating the end points of each subinterval
in the time-sliced expressions.
Moreover, the perturbation
expansions based on 
Eqs.~(\ref{eq:propagator_perturb_expansion})--(\ref{eq:GreenF_perturb_expansion_nth_order})
admit the usual diagrammatic representations.

\subsection{Central Potentials}
\label{sec:central_potentials}

For the case of central potentials,
the same procedure can be directly applied to the radial path integrals,
with their respective Besselian
pseudomeasures~(\ref{eq:radial_pseudomeasure}) at each order.
Alternatively, 
Eqs.~(\ref{eq:propagator_perturb_expansion})--(\ref{eq:GreenF_perturb_expansion_nth_order})
 could be resolved in hyperspherical coordinates term by term---as in the derivation
leading to Eq.~(\ref{eq:propagator_partial_wave_exp}). 
Either derivation
yields the exact infinite summation
\begin{equation}
K_{l+\nu} (r'',r';T) =
\sum_{n=0}^{\infty}
K^{(n)}_{l+\nu} (r'',r';T) 
\;  ,
\label{eq:propagator_perturb_expansion_radial}
\end{equation}
for the propagator, where 
\begin{eqnarray}
K^{(n)}_{l+\nu} (r'',r';T) 
& = &
\prod_{\alpha=1}^{n} \left[
 \int_{0}^{\infty} 
d r_{\alpha} \,
\frac{   V(r_{\alpha})  }{i \hbar} 
\right]
\nonumber 
\\
& \times &
\int_{t'}^{t''} dt_{n} 
\int_{t'}^{t_{n}} dt_{n-1} 
\ldots
\int_{t'}^{t_{2}} dt_{1} 
\,
\prod_{\beta=0}^{n} 
\left[
K^{(0)}_{l+\nu} (r_{\beta+1},r_{\beta};
t_{\beta+1}-t_{\beta} )  \right]
\;   ,
\label{eq:propagator_perturb_expansion_radial_nth_order}
\end{eqnarray}
Correspondingly, the 
energy Green's function associated
with Eqs.~(\ref{eq:propagator_perturb_expansion_radial}) and 
(\ref{eq:propagator_perturb_expansion_radial_nth_order}) is
\begin{equation}
G_{l+\nu} (r'',r';E) =
\sum_{n=0}^{\infty}
G^{(n)}_{l+\nu} (r'',r';E) 
\; ,
\label{eq:GreenF_perturb_expansion_radial}
\end{equation}
where
\begin{equation}
G^{(n)}_{l+\nu} (r'',r';E) 
=
\prod_{\alpha=1}^{n} 
\left[ \int_{0}^{\infty}
d r_{\alpha} 
 V(r_{\alpha})  \right]
\,
\prod_{\beta=0}^{n} \left[ 
G^{(0)}_{l+\nu} (r_{\beta +1},r_{\beta};E) 
  \right]
\;  .
\label{eq:GreenF_perturb_expansion_radial_nth_order}
\end{equation}
Finally, in subsequent applications it will prove convenient to introduce the reduced
functions ${\cal V} (r)$
 and $ {\mathcal G}^{(n)}_{l+\nu} (r'',r';E)    $,
implicitly defined from
\begin{equation}
V ( r ) = - \frac{\hbar^{2}}{2M}\, \lambda \,
{\mathcal V} (r)
\;  
\label{eq:reduced_potential}
\end{equation}
and
\begin{equation}
G^{(n)}_{l+\nu} (r'',r';E)    
 = 
 -
\frac{2M}{\hbar^{2}} 
\,
\left| 
{\mathcal V} (r'')
{\mathcal V} (r')
\right|^{-1/2}
\,
{\mathcal G}^{(n)}_{l+\nu} (r'',r';E)    
\;  ,
\label{eq:reduced_GreenF}
\end{equation}
as well as the sign
symbol
$\epsilon_{r_{\alpha}} = {\rm sgn} \, (r_{\alpha})$.
These substitutions
lead to the reduced perturbation expansion
\begin{equation}
{\mathcal G}_{l+\nu} (r'',r';E) =
\sum_{n=0}^{\infty}
{\mathcal G}^{(n)}_{l+\nu} (r'',r';E) 
\; ,
\label{eq:reduced_GreenF_perturb_expansion_radial}
\end{equation}
with
\begin{equation}
{\mathcal G}^{(n)}_{l+\nu} (r'',r';E) 
=
\lambda^{n}
\,
\prod_{\alpha=1}^{n} 
\left[ 
\int_{0}^{\infty}
d r_{\alpha}  \,  \epsilon_{r_{\alpha}}
\right]
\,
\prod_{\beta=0}^{n} \left[ 
{\mathcal G}^{(0)}_{l+\nu} (r_{\beta +1},r_{\beta};E) 
  \right]
\;  .
\label{eq:reduced_GreenF_perturb_expansion_radial_nth_order}
\end{equation}
Notice that, even for $n=0$ in Eq.~(\ref{eq:reduced_GreenF}),
the ``unperturbed'' function
${\mathcal G}^{(0)}_{l+\nu} (r'',r';E) $ 
already includes information about the perturbation
${\mathcal V} (r)$.

\subsection{Integral Representations}
\label{sec:integral_representations}

The perturbation-theory framework of the previous subsection can be further extended
when
${\mathcal V} (r)$ is a monotonic function. This is precisely
the case for the class of power-law 
interactions---including the inverse square potential. 
When this condition is satisfied,
the sign symbols
$\epsilon_{r_{\alpha}}$ in
Eq.~(\ref{eq:reduced_GreenF_perturb_expansion_radial_nth_order})
can be omitted altogether.
As a result,
the perturbation expansion of Eqs.~(\ref{eq:reduced_GreenF_perturb_expansion_radial})
and (\ref{eq:reduced_GreenF_perturb_expansion_radial_nth_order})
reduces to the evaluation of
\begin{eqnarray}
{\mathcal G}^{(n)}_{l+\nu} (r'',r';E) 
&
=
&
\lambda^{n}
\,
\int_{0}^{\infty}
  d r_{n} 
\,
\ldots
\,
\int_{0}^{\infty}
 d r_{1}  
\,
\nonumber \\
& \times &
{\mathcal G}^{(0)}_{l+\nu} (r'',r_{n};E) 
{\mathcal G}^{(0)}_{l+\nu} (r_{n},r_{n-1};E) 
\ldots
{\mathcal G}^{(0)}_{l+\nu} (r_{2},r_{1};E) 
{\mathcal G}^{(0)}_{l+\nu} (r_{1},r';E) 
\;  .
\label{eq:reduced_GreenF_perturb_expansion_radial_nth_order_ISP}
\end{eqnarray}
The formal structure
of Eq.~(\ref{eq:reduced_GreenF_perturb_expansion_radial_nth_order_ISP})
can be interpreted to be of the operator form
\begin{equation}
\hat{ {\mathcal G} }^{(n)}_{l+\nu} (E) 
= 
\lambda^{n}
\, 
\left[ 
\hat{\mathcal G}^{(0)}_{l+\nu} (E)
\right]^{n+1}
\; ,
\label{eq:reduced_GreenF_radial_nth_order_ISP_operator}
\end{equation}
according to the prescriptions presented in the Appendix,
with operator products evaluated in real space.
This observation leads to a reconstruction of the operator 
$\hat{ {\mathcal G} }^{(n)}_{l+\nu} (E) $ in terms of the eigenvalues of
$\hat{\mathcal G}^{(0)}_{l+\nu} (E)$, as the following argument shows.

First, suppose that the eigenvalues and eigenvectors
of the self-adjoint operator 
$\hat{\mathcal G}^{(0)}_{l+\nu} (E)$ are
labeled with an index $z$ and
known to be $\eta_{l+\nu}(z;\kappa)$ and
$\Psi_{z,l+\nu}(\kappa)$ respectively,
where $\kappa$ is given by Eq.~(\ref{eq:imaginary_wave_number}). 
In symbolic notation,
\begin{equation}
\hat{ {\mathcal G}}^{(0)}_{l+\nu} (E)
\,
\Psi_{z,l+\nu}(\kappa)
=
\eta_{l+\nu} (z;\kappa)
\,
\Psi_{z,l+\nu}(\kappa)
\;   ,
\label{eq:eigenvalue_problem_G_zero}
\end{equation} 
which stands for the integral relation
\begin{equation}
\int_{0}^{\infty}
  d r'  
\,
{\mathcal G}^{(0)}_{l+\nu} (r,r';E)
\Psi_{z,l+\nu} (r';\kappa)
=
\eta_{l+\nu}(z;\kappa)
\Psi_{z,l+\nu} (r;\kappa)
\;  .
\label{eq:eigenvalue_problem_G_zero_integral}
\end{equation}

Second, in operator symbolic notation,
Eq.~(\ref{eq:reduced_GreenF_radial_nth_order_ISP_operator})
is equivalent to the eigenvalue equation
\begin{equation}
\hat{ {\mathcal G}}^{(n)}_{l+\nu} (E)
\,
\Psi_{z,l+\nu}(\kappa)
=
\lambda^{n}
\, 
\left[ 
\eta_{l+\nu} (z;\kappa)
\right]^{n+1}
\,
\Psi_{z,l+\nu}(\kappa)
\;   .
\label{eq:reduced_GreenF_radial_nth_order_ISP_eigenvalue_eq}
\end{equation} 

Third,
the corresponding eigenvalue equation
for the radial Green's 
function operator 
$\hat{ {\mathcal G}}_{l+\nu} (E)$ 
can be obtained by geometrically 
{\em summing the entire perturbation series\/},
\begin{equation}
\hat{ {\mathcal G}}_{l+\nu} (E)
\,
\Psi_{z,l+\nu} (\kappa)
=
\frac{1}{  \left[ \eta_{l+\nu}(z;\kappa) \right]^{-1}
 -  \lambda}
\;
\Psi_{z,l+\nu} (\kappa)
\;   .
\label{eq:operator_G}
\end{equation} 
This equation defines implicitly 
the operator $\hat{ {\mathcal G}}_{l+\nu} (E)$.

Finally, 
Eq.~(\ref{eq:operator_G}) can be recast into 
the explicit form
\begin{equation}
{\mathcal G}_{l+\nu} (r'',r';E) 
=
\int_{0}^{\infty}  
dz
\,
\frac{1}{  \left[ \eta_{l+\nu}(z;\kappa) \right]^{-1}
 -  \lambda}
\;
\Psi_{z,l+\nu} (r'';\kappa) \Psi_{z,l+\nu} (r';\kappa)
\;  ,
\label{eq:GreenF_spectral_reps}
\end{equation} 
which stands for the Green's spectral representation
(see Appendix).
It should be noticed that 
the energy dependence of 
${\mathcal G}_{l+\nu} (r'',r';E) $
arises from possibly both
the eigenfunctions $\Psi_{z,l+\nu} (r;\kappa)$ and eigenvalues
$\eta_{l+\nu}(z;\kappa) $.

The procedure outlined in the reconstruction of the previous paragraphs
illustrates how the formal infinite summation can be implemented.
However, a caveat is in order:
Eq.~(\ref{eq:GreenF_spectral_reps}) 
can only be turned into an explicit solution
if the eigenvalue problem
of Eq.~(\ref{eq:eigenvalue_problem_G_zero}) is solved first.
As an alternative,
the solution to Eq.~(\ref{eq:eigenvalue_problem_G_zero}) 
could be directly identified if a corresponding spectral representation
\begin{equation}
 {\mathcal G}^{(0)}_{l+\nu} 
(r'',r';E) 
=
\int_{0}^{\infty}  
dz
\,
\eta_{l+\nu}(z;\kappa)
\,
\Psi_{z,l+\nu} (r'';\kappa) \Psi_{z,l+\nu} (r';\kappa)
\;   .
\label{eq:GreenF_order_zero_spectral_reps}
\end{equation} 
were found.
In short, 
the existence of an integral representation~(\ref{eq:GreenF_order_zero_spectral_reps})
implies directly the form~(\ref{eq:GreenF_spectral_reps})
for the exact Green's function 
$ {\mathcal G}_{l+\nu} 
(r'',r';E) $.

For the remainder of this paper we will explore particular applications 
of this remarkable summation technique.

\section{Quantum-Mechanical Propagator for the Inverse Square Potential}
\label{sec:ISP}

In this section we start a path-integral analysis of the
inverse square potential 
\begin{equation}
V ( r ) = - \frac{\hbar^{2}}{2M}\, \frac{\lambda}{ r^{2} } 
\;  ,
\label{eq:ISP}
\end{equation}
which we conveniently rewrite in such a way that 
${\mathcal V}= 1/r^{2}$ [cf.\ Eq.~(\ref{eq:reduced_potential})]
and
$\lambda$ be dimensionless; 
moreover, $\lambda <0$ describes a repulsive potential, while 
$\lambda >0$ describes an attractive interaction.
Direct evaluation of Eq.~(\ref{eq:propagator_QM_spherical_coords_2D_expansion}) 
leads to complicated integrals and recursion relations that cannot be easily tamed
(unlike the cases of a free particle and a harmonic oscillator). Instead,
a judicious application of the method of infinite summation of perturbation theory
leads to an exact analytical solution, which will constitute the main result of this section.

Before considering the 
complete solution, 
one may notice that 
Eqs.~(\ref{eq:propagator_free_particle}) and (\ref{eq:GreenF_free_particle})
describe physics of a free particle, for each angular momentum channel,
and thereby suggest the following straightforward heuristic argument.
First,   
one may consider 
  the asymptotic
behavior of the Bessel function,
\begin{equation} 
\sqrt{2 \pi z} \,
e^{-z}
\,
 I_{p}(z) \stackrel{(z \rightarrow \infty)}{\sim} 
1 - \frac{(p^{2} -1/4)}{2z}
\; ,
\label{eq:Bessel_asymptotics}
\end{equation}
in the limit
$N \rightarrow \infty$,
which according to Eq.~(\ref{eq:z_variable}) amounts to $z \rightarrow \infty$.
Second, 
the asymptotics~(\ref{eq:Bessel_asymptotics}) can be applied 
to each factor in the pseudomeasure~(\ref{eq:radial_pseudomeasure}).
Third, each exponential factor can be combined with and compared against a corresponding 
factor arising from the inverse square interaction.
In conclusion,
this
shows that the only effect of adding an inverse square potential~(\ref{eq:ISP}) 
to the propagator~(\ref{eq:propagator_QM_spherical_coords_2D_expansion})
is a shift in the angular momentum quantum number,
\begin{equation}
l+ \nu \rightarrow s_{l}
=\sqrt{ (l+\nu)^{2} - \lambda}
\;  .
\label{eq:free_particle_ISP_replacement}
\end{equation}
Even though this replacement is straightforward in the 
Schr\"{o}dinger formulation of the theory,
its justification in the path-integral approach 
is far from obvious. Not surprisingly,
the asymptotic argument has been called into 
question~\cite{kleinert}. Thus, 
 we will first refine the argument above and 
show how to confirm this conjecture~(\ref{eq:free_particle_ISP_replacement})
in a more satisfactory way.

\subsection{Integral Representation of the Green's Function for the Inverse Square Potential} 
\label{sec:integral_representation_ISP}

The derivation of the result of Eq.~(\ref{eq:free_particle_ISP_replacement})
can be completed in a number of equivalent ways. The 
most elegant route is based on direct transformations at the level of the
 energy Green's functions. 
For the inverse square potential~(\ref{eq:ISP}),
the perturbation expansion of Eqs.~(\ref{eq:reduced_GreenF_perturb_expansion_radial})
and (\ref{eq:reduced_GreenF_perturb_expansion_radial_nth_order})
reduces to the evaluation of 
Eqs.~(\ref{eq:reduced_GreenF_perturb_expansion_radial_nth_order_ISP})--(\ref{eq:GreenF_spectral_reps}).

Remarkably,
an integral representation of the 
form~(\ref{eq:GreenF_order_zero_spectral_reps})
is known, as first pointed in Ref.~\cite{bha:89};
explicitly,
it is the Kontorovich-Lebedev representation~\cite{kontorovich}  
\begin{equation}
G^{(0)}_{l+\nu} (r'',r';E)    
=
 - \frac{2M}{\hbar^{2}} \,
\sqrt{ r'r''} \,
{\cal J}_{l+\nu} (\kappa r'',\kappa r';1)
\;   ,
\label{eq:GreenF_free_particle_KL}
\end{equation}
where
\begin{equation}
{\cal J}_{p} (\xi,\eta; \alpha)
=
\frac{2}{\pi^{2}}
\,
\int_{0}^{\infty}  
dz \; \frac{ z \sinh ( \pi z )}{ \left(z^{2} + p^{2} \right)^{\alpha} }
\,
K_{iz} (\xi) K_{iz} (\eta)
\;   .
\label{KL_integral}
\end{equation}
In Eq.~(\ref{eq:GreenF_free_particle_KL}) 
the generalized function
$
{\cal J}_{p} (\xi,\eta; \alpha)
$
reduces to the product 
of Bessel functions in 
Eq.~(\ref{eq:GreenF_free_particle}),
\begin{equation}
{\cal J}_{p} (\xi,\eta; 1)
=
   \sqrt{r'r''}  \,
 {\mathcal G}^{(0)}_{l+\nu} (r,r';E) 
=
I_{l+\nu}( \kappa r_{<} ) K_{l+ \nu} (\kappa r_{>} )
\;  ,
\end{equation}
 when $\alpha=1$.

In the Kontorovich-Lebedev representation~\cite{kontorovich},
which we review in the Appendix,
the Dirac-normalized complete set of functions
$
\Psi_{z} (r;\kappa)
=
\sqrt{  2 z \sinh ( \pi z ) /(\pi^{2}  r )  }
\;
K_{iz} (\kappa r) $
permits the identification of 
Eq.~(\ref{KL_integral})  
[supplemented with~Eq.~(\ref{eq:GreenF_free_particle_KL})]
with Eq.~(\ref{eq:GreenF_spectral_reps}). 
Thus,
Eq.~(\ref{eq:eigenvalue_problem_G_zero}) 
is realized with
\begin{equation}
 \eta_{l+\nu}(z)
 = 
\frac{1}{z^{2} + (l+\nu)^{2} }
\;  ,
\label{eq:GreenF_order_zero_eigenvalues}
\end{equation}
so that the eigenvalues~(\ref{eq:operator_G}) of 
$\hat{ {\mathcal G}}_{l+\nu} (E)$
are 
\begin{equation}
\frac{1}{
\left[ \eta_{l+\nu}(z) \right]^{-1}
- \lambda }
= 
\frac{1}{
z^{2} + s_{l}^{2}
}
\;  .
\label{eq:GreenF_eigenvalues}
\end{equation}
It should be noticed that
$ \eta_{l+\nu}(z)$ is independent of the energy, while
$\Psi_{z}$ is independent of $(l+\nu)$;
this is characteristic of the  Kontorovich-Lebedev representation.
Finally,
the remarkable reconstruction~(\ref{eq:GreenF_spectral_reps}) can be explicitly carried out,
\begin{eqnarray}
G_{l+\nu} (r'',r';E| \lambda) 
& = &
 - 
 \frac{2M}{\hbar^{2}} \,
\sqrt{ r'r''} \,
\sum_{n=0}^{\infty}
\lambda^{n}
{\cal J}_{l+\nu} (\kappa r'',\kappa r'; n + 1)
\nonumber \\
& = &
 - \frac{2}{\pi^{2}}
 \frac{2M}{\hbar^{2}} \,
\int_{0}^{\infty}  dz \, 
\frac{ z \sinh ( \pi z )}{  z^{2} + s_{l}^{2} }
\,
K_{iz} (\kappa r'') K_{iz} (\kappa r')
\label{eq:GreenF_KL}
\\
& = &
-
 \frac{2M}{\hbar^{2}} \,
\sqrt{ r'r''} \,
{\cal J}_{s_{l}} (\kappa r'',\kappa r'; 1)
\;     .
\label{eq:GreenF_ISP}
\end{eqnarray}
Equation~(\ref{eq:GreenF_ISP}) 
is identical in form with~(\ref{eq:GreenF_free_particle_KL}); thus, the
replacement~(\ref{eq:free_particle_ISP_replacement})
is now proved  within the path-integral formulation.
 In short, going back to Eq.~(\ref{eq:GreenF_free_particle}), 
\begin{equation}
G_{l+\nu} (r'',r';E| \lambda) 
=  G^{(0)}_{s_{l}} (r'',r';E) 
=
- \frac{2M}{\hbar^{2}} \,
\sqrt{ r'r''} \,
I_{s_{l}}( \kappa r_{<} ) K_{s_{l}} (\kappa r_{>} )
\; ,
\label{eq:GreenF_ISP2}
\end{equation}
while its
inverse Fourier transform 
is identical to $K^{(0)}_{s_{l}} (r'',r';T)  $, so that
[cf.~(\ref{eq:propagator_free_particle})]
\begin{equation}
K_{l+\nu} (r'',r';T | \lambda) 
=
\frac{M}{i \hbar T} 
\,
\sqrt{ r' r''} 
\,
\exp \left[ \frac{i M}{2 \hbar T} \left(r^{\prime 2} + r^{\prime \prime 2} 
\right) \right]
I_{s_{l}} \left( \frac{Mr'r''}{i \hbar T} \right)
\; .
\label{eq:propagator_ISP}
\end{equation}

\subsection{Nature of the Solution}
\label{sec:nature_of_solution}

There is only one apparent limitation in the above derivation:
the geometric series involved in the infinite perturbation 
expansion is guaranteed to converge only
for $|\lambda| < \lambda_{l}^{(\ast)} $, where 
$\lambda_{l}^{(\ast)} =(l+\nu)^{2}$. 
This condition
has been regarded as an actual limitation~\cite{bha:89};
however, as a restriction, it can be 
immediately lifted by noticing that the final 
expression~(\ref{eq:GreenF_ISP2}) 
provides the desired analytic continuation in the complex $\lambda$ plane.

Moreover, for each angular momentum state,
the radius of convergence of the above series is set by the 
existence of a critical point 
$\lambda = \lambda_{l}^{(\ast)} $, at which the
nature of the path integral for real $\lambda$ changes.
In fact, the integral representation~(\ref{eq:GreenF_KL})
explicitly displays that there exist two fundamentally distinct regimes
with respect to the coupling parameter $\lambda$ (dependence which
is encoded in $s_{l}$):
\begin{itemize}
\item
{\sf 
Weak-coupling regime\/}:
$ \lambda <\lambda_{l}^{(\ast)}$, including repulsive potentials.

The order $s_{l}$ of the Bessel functions  is real
and Eq.~(\ref{eq:GreenF_ISP2}) has no singularities for real $\kappa$, 
showing that
the system cannot sustain bound states---a manifestation of 
the scale invariance of the potential $r^{-2}$.

\item
{\sf 
Strong-coupling regime\/}:
$ \lambda >\lambda_{l}^{(\ast)} $.

The Bessel functions acquire an imaginary order $s_{l}=i \Theta_{l}$,
where 
\begin{equation}
\Theta_{l} 
= [ \lambda - \lambda_{l}^{(\ast)} ]^{1/2}
\;  ,
\label{eq:theta_coupling}
\end{equation}
 and the negative-energy states form a
continuous spectrum not bounded from below. 
\end{itemize}

 The pathology displayed in the strong-coupling
spectrum can be avoided by the use of field-theory
regularization and renormalization techniques,
as shown in Refs.~\cite{gup:93,cam:00,cam:01}.
Accordingly, introducing a short-distance cutoff
$a$, 
the resulting radial Green's function
$ G_{l+\nu} (r'',r';E|\lambda;a)$
inherits the
regular boundary condition 
\begin{equation}
\left. G_{l+\nu} (r'',r';E|\lambda;a)
\right|_{r''=a \, {\rm or } \, r'=a}  
=0
\;  .
\label{eq:regularized_BC}
\end{equation} 

However, the implementation of
Eq.~(\ref{eq:regularized_BC}) is not straightforward in the path-integral formalism. 
The main difficulty encountered in a path-integral real-space 
renormalization with the Dirichlet 
boundary condition~(\ref{eq:regularized_BC})
lies in the proper implementation of the sum over all paths.
A technique for dealing with this difficulty,
which was originally introduced in 
Ref.~\cite{gro:93}, amounts to adding an infinite-strength
repulsive delta-function potential $\sigma \delta (r-a)$.
Therefore, we define the 
regularized radial energy Green's function 
$G_{l+\nu} (r'', r' ; E| \lambda; a)$
as the limit
$\sigma \rightarrow \infty$ of the 
more general Green's function $G_{l+\nu} (r'', r' ; E| \lambda; a; \sigma)$
in the presence of the delta-function perturbation.
We now turn our attention to this problem.

\section{Inverse Square Potential in the Presence of a  
Delta-Function Interaction}
\label{sec:ISP+delta}

\subsection{Computation of the Green's Function}
\label{sec:ISP+delta_GF_computation}

We will now compute the
general Green's function $G_{l+\nu} (r'', r' ; E| \lambda; a; \sigma)$
in the presence of an interaction
\begin{equation}
V_{\rm total}(r)=
-\frac{ \hbar^{2}}{2M}
 \,  \frac{\lambda}{r^{2}}
+ \sigma \delta (r-a)
\,  .
\label{eq:ISP+delta}
\end{equation}
This problem can be most effectively analyzed by 
treating it exactly to all orders
as a delta-function perturbation 
to an inverse square potential.
More precisely,
the unperturbed action
$S^{(0)}$  includes both the kinetic term and the inverse-square interaction,
with their exact physics described by 
$G_{l+\nu} (r'',r';E| \lambda)$, 
Eq.~(\ref{eq:GreenF_ISP2}),
or alternatively by 
$K_{l+\nu} (r'',r';T | \lambda)$, Eq.~(\ref{eq:propagator_ISP}).
In addition, the 
delta-function term 
$V (r)= \sigma \delta (r-a)$
is to be regarded
as a perturbation to which the theory of Secs.~\ref{sec:general_framework}
and \ref{sec:infinite_summation}
applies.
 
Then, going back to Eq.~(\ref{eq:GreenF_perturb_expansion_radial_nth_order}),
each factor
$\int_{0}^{\infty}
d r_{\alpha} \,
 V( r_{\alpha})  
$
is merely reduced to $\sigma$ and carries the additional instruction that
the replacement 
$r_{\alpha} \rightarrow a$ be made;
then,
\begin{equation}
G^{(n)}_{l+\nu} 
\left(
\left. 
r'', r' ; E  
\right| \lambda; a; \sigma
\right)
=
\sigma^{n}
\,
\left[
G_{l+\nu}(a,a;E|\lambda) 
\right]^{n-1}
\,
G_{l+\nu} (r'', a ; E| \lambda)
\,
G_{l+\nu} ( a, r' ; E| \lambda)
\;  
\label{eq:reduced_GreenF_delta_perturb_expansion_radial_nth_order}
\end{equation}
for $n \geq 1$, while the term of order zero distinctly remains
$G_{l+\nu} (r'', r' ; E| \lambda) $.
The exact infinite summation of this series 
for finite $\sigma$ leads to the familiar result~\cite{gro:93}
\begin{equation}
G_{l+\nu} 
\left(
\left. 
r'', r' ; E  
\right| \lambda; a; \sigma
\right)
=
G_{l+\nu} (r'', r' ; E| \lambda)
 -
\frac{ G_{l+\nu} (r'', a ; E| \lambda)
\, G_{l+\nu} ( a, r' ; E| \lambda)
}{
G_{l+\nu}(a,a;E|\lambda) 
- 1/\sigma }
\;  .
\label{eq:PI_Dirichlet_BC}
\end{equation}
It should be noticed that 
Eq.~(\ref{eq:PI_Dirichlet_BC})
describes the complete physics of a delta-function perturbation to {\em any\/}
known problem with action  $S^{(0)}$ and
 exactly described by $G_{l+\nu} (r'', r' ; E| \lambda)$.

\subsection{Applications}
\label{sec:applications}

The  combined interaction~(\ref{eq:ISP+delta})
is of current interest, above and beyond its application
to the regularization of the inverse square potential.
Recent work in M-theory has led to a particular
realization of this combined potential~\cite{Randall-Sundrum}
and the use of Green's functions was advanced for further analysis~\cite{Park},
within a more general investigation of gravitation in our
familiar four-dimensional world as arising from a higher-dimensional theory. 
Even though this example corresponds to a 
very specific model, it shows the usefulness of the examination
of singular potentials with the techniques introduced in this paper.

Let us now briefly review the relevant example,
which arises in the description of
effective four-dimensional 
gravitational effects observed on a $(3 + 1)$-dimensional 
subspace, a 
3-brane,
embedded in a spacetime
with five noncompact dimensions ($AdS_{5}$)~\cite{Randall-Sundrum}.
The
linearized tensor fluctuations of the metric 
are subject to a Kaluza-Klein reduction
and assumed to have a dependence $h(x,y) = \psi (y) e^{p \cdot x}$
in terms of the coordinates $x$ on the brane
and the extra-dimensional coordinate $y$.
The problem is then reduced to an {\em effective\/} one-dimensional
Schr\"{o}dinger equation with respect to the 
``wave function'' $\psi (y)$ in terms of the extra-dimensional coordinate.
 Even though
 direct reduction leads to a complicated potential with respect to $y$,
the transformation of variables
 $z=  {\rm sgn} \;  (y) \times (e^{k|y|} -1)/k$,
$\hat{\psi} (z) = \psi (y) e^{k|y|/2}$
simplifies the effective 
Schr\"{o}dinger problem to
($\hbar=1$, $M=1$)
\begin{equation}
\left\{
-\frac{1}{2} \partial^{2}_{z} +
\left[
 - \frac{\lambda}{ 2 (|z| + a)^{2}}
+ \sigma \delta (z)
\right]
\right\}
\hat{\psi} (z)
=
m^{2}
\, \hat{\psi} (z)
\; ,
\label{eq:ISP+delta_applied_to_branes}
\end{equation}
with the following specific parameters:
\begin{eqnarray}
\lambda = - \frac{15}{4}
\nonumber \\
\sigma =- 3k
\nonumber \\
a= \frac{1}{k}
\; ,
\label{eq:brane_parameters}
\end{eqnarray}
as dictated by the physics of the Kaluza-Klein reduction.
More precisely,
the additional coordinate change 
$\xi = |z| + a$, shows that
Eq.~(\ref{eq:ISP+delta_applied_to_branes}) indeed 
describes an effective interaction of the form~(\ref{eq:ISP+delta})
with respect to the coordinate $\xi$.
It should be noticed 
that the signs
in this problem (attractive delta and repulsive inverse square potential)
are just the opposite of the 
those needed for the regularization of the single inverse square potential, 
as discussed in the next section.

 The effective dimensionality $D=1$  ($\nu= -1/2$)
of problem~(\ref{eq:ISP+delta_applied_to_branes})
implies that the only available channels are $l=0$ and $l=1$,
so that $l+\nu = \mp 1/2$ respectively;
from this
 and Eqs.~(\ref{eq:free_particle_ISP_replacement}) and (\ref{eq:brane_parameters}),
the corresponding order of the Bessel functions is 
 $s_{l} = 2$. 
Even though
the repulsive inverse square potential is incapable
of producing bound states by itself,
the additional presence of a one-dimensional attractive delta-function perturbation
yields exactly one bound state~\cite{Randall-Sundrum}.
 This is interpreted as corresponding to a massless
four-dimensional graviton---a fact that has been used to confirm the claim that 
the experimentally observed four-dimensional gravitational fields can arise through 
a hypothetical scenario of dimensional
reduction from extra noncompact dimensions.
Additional details on this problem and related applications are currently 
under investigation.

Parenthetically, a related and simple example
of the application of these
techniques is afforded by
the two-dimensional delta-function potential 
$V({\bf r}) = - 
\hbar^{2} g
\,
\delta^{(2)} ({\bf r})/2M$, for which the
propagator is also singular and calls for regularization. 
For example, using
a real-space regulator $a$, one may replace
   $\delta^{(2)}({\bf r})$ by
$ \delta (r-a)/2 \pi a$.
Then, selecting
$\lambda=0$, $\nu=0$, and $\sigma= - \hbar^{2} g/2M$ in
Eq.~(\ref{eq:PI_Dirichlet_BC}) and identifying the pole(s), 
the regularized equation for the bound state in the $s$ channel
becomes
$
 K_{0} (\kappa a) = 2 \pi/g
$,
whose renormalized counterpart agrees with the known 
answer~\cite{delta},
as we have recently shown~\cite{cam:01c}.

\section{Regularization of the Inverse Square Potential}
\label{sec:ISP_regularization}

We now turn our attention to the problem that was anticipated in Sec.~\ref{sec:ISP}:
the real-space regularization of an inverse square potential with supercritical coupling.
In this procedure, a short-distance regulator $a$ is introduced and the
boundary condition~(\ref{eq:regularized_BC}) is enforced
as the
 $\sigma \rightarrow \infty$ limit
of the 
Green's function $G_{l+\nu} (r'', r' ; E| \lambda; a; \sigma)$
in the presence of the delta-function perturbation---the second term
in Eq.~(\ref{eq:ISP+delta}).

Therefore,  
Eqs.~(\ref{eq:GreenF_ISP2}) and (\ref{eq:PI_Dirichlet_BC})
provide the desired regularized radial energy Green's function 
\begin{equation}
G_{l+\nu} 
\left(
\left. 
r'', r' ;  
E
\right| 
\lambda; a
\right)
=
 -  \frac{2M}{\hbar^{2}} \,
\frac{\sqrt{ r'r''}}{ K_{s_{l}}(\kappa a)}
\,
\left[
K_{s_{l}}(\kappa a) I_{s_{l}}(\kappa r_{<})
-
I_{s_{l}}(\kappa a) K_{s_{l}}(\kappa r_{<})
\right]
K_{s_{l}}(\kappa r_{>})
\;  ,
\label{eq:GreenF_l_reg}
\end{equation}
where $\kappa=
\sqrt{-2ME}/\hbar$ 
[cf. Eq.~(\ref{eq:GreenF_free_particle})].
Equation~(\ref{eq:GreenF_l_reg}) is in perfect agreement with
the result from 
the operator approach to a Green's function: 
$G(r'',r';E) = 
 u_{<} (r_{<}) u_{>} (r_{>})/ p W[u_{<},u_{>}]$,
where $p= \hbar^{2}/2M$, while $u_{<}(r)$ and $u_{>}(r)$ are
the solutions satisfying the boundary condition at $r=a$
and at infinity~\cite{Green_operator_approach}.

The poles of Eq.~(\ref{eq:GreenF_l_reg}), which 
are implicitly given by
\begin{equation}
K_{s_{l}} (\kappa a) = 0
\;  ,
\label{eq:reg_eigenvalue_eq}
\end{equation}
yield the bound-state sector of the theory.
The energy levels
can be derived by specializing to the case when $a$ is small, 
in which case the small-argument expansion of the modified Bessel function
of the second kind becomes~\cite{cam:00,abr:72}
\begin{equation}
K_{i\Theta_{l}} (z) 
  \stackrel{(z \rightarrow 0)}{\sim}
-
\sqrt{ \frac{\pi}{ \Theta_{l} 
\sinh  \left( \pi \Theta_{l} \right) } }
\,
\sin
\left[
\Theta_{l} \ln \left( \frac{ z }{2} \right) 
- \delta_{\Theta_{l}}
\right]
\,
\left[ 1 + O \left( z^{2} \right)  \right] 
\;  ,
\label{eq:ISP_weak_BS_asymptotic_wf}
\end{equation}
in which
$ \delta_{\Theta_{l}} $ is the phase of
$\Gamma (1+i\Theta_{l})$.
In particular,
Eq.~(\ref{eq:ISP_weak_BS_asymptotic_wf})
has an infinite set of zeros when the order of the 
 modified Bessel function is imaginary; these zeros
are
\begin{equation}
z_{n} = 2 \,
e^{ (\delta_{\Theta_{l}} - n \pi)/\Theta_{l} }
\, 
\end{equation}
[up to a correction factor  $1 + O(z_{n}^{2}/\Theta_{l} )$],
where $n$ is an integer. Furthermore,
with the assumption that $z_{n} \ll 1$ and
$\Theta_{l} \geq 0$,  it follows that $(-n)<0$,
whence  $n= 1, 2, 3, \ldots$.
Parenthetically, $z_{n} \ll 1$ only if $\Theta_{l} \ll 1$, so that
$\delta_{\Theta_{l}}=- \gamma \Theta_{l}+ O(\Theta_{l}^{2})$
(with $\gamma$ being the Euler-Mascheroni constant).
This argument shows that
the
energy levels are given by
\begin{equation}
E_{ n_{r} l }
=
-
\frac{2 \hbar^{2} e^{-\gamma}}{M a^{2} }
\,
\exp 
\left( 
- \frac{2 \pi n_{r} }{ \Theta_{l} } 
\right)
\; ,
\end{equation}
in which 
$n=n_{r}$ is the usual radial quantum number.

The regularization of 
Eq.~(\ref{eq:reg_eigenvalue_eq})
provides the foundation for the next step: renormalization.
This final step may be implemented by
demanding the dependence of the coupling with respect to the
regulator; from Eq.~(\ref{eq:theta_coupling})
this implies the dependence
 $\Theta_{l}=\Theta_{l} (a)$
in the limit $a \rightarrow 0$~\cite{running_coupling}, 
as in Refs.~\cite{gup:93,cam:00,cam:01}. 
For example, when this procedure is applied to the regularized ground state,
the required relation becomes
\begin{equation}
- 
g^{(0)}
=
\frac{2 \, \pi}{ \Theta_{_{\rm (gs)}} (a) } +
2 \, \ln \left( \frac{\mu \, a}{2} \right)
+ 2 \, \gamma
\;  ,
\label{eq:cutoff_BS_regularized_relation}
\end{equation}
where $\mu$ is an arbitrary
renormalization scale with dimensions of inverse length
and $g^{(0)}$ is an arbitrary finite part associated with the
coupling, such that
\begin{equation}
E_{_{\rm (gs)}}
=
- 
\frac{\hbar^{2}
\mu^{2}}{2M}
\,
\exp \left[
g^{(0)} 
\right]
\;  .
\label{eq:ISP_GS}
\end{equation}
The ground-state wave function 
is obtained in the limit
$\Theta_{_{\rm (gs)}}    (a) 
 \stackrel{(a \rightarrow 0 )}{\longrightarrow} 
0^{+}$, so that,
from Eqs.~(\ref{eq:residue_magnitude_wf})
and (\ref{eq:residue_phase_wf}),
\begin{equation}
\Psi_{_{\rm (gs)}}
 ({\bf r})
=
\sqrt{
\Gamma \left( \frac{D}{2} \right) \,
\left( 
\frac{\mu^{2}}{\pi}
\right)^{D/2}
}
\;
\frac{K_{0} (\mu r)}
{ \left( \mu r \right)^{D/2 -1} }
\;  .
\label{eq:ISP_wf_normalized_renormalized}
\end{equation}

The renormalization
procedure described above leads to the emergence of an arbitrary 
dimensional scale---phenomenon known as dimensional transmutation---and 
violates the manifest SO(2,1) invariance of the 
theory~\cite{jac:72,alf:76,jac:80,cam:01b}. This is
an instance of an established 
quantum anomaly, which manifests itself in the
three-dimensional molecular realm 
for the interaction between polar molecules and electrons~\cite{cam:01b}.

Finally, the scattering sector of the theory can be analyzed in a similar way,
with the Bessel functions in Eq.~(\ref{eq:GreenF_l_reg})
replaced as follows:
$I_{s} (\kappa r) 
\rightarrow  
(-i)^{s}
J_{s} (k r) $
and
$K_{s} (\kappa r) 
\rightarrow 
\pi i^{s+1}H^{(1)}_{s} (k r)/2 $, 
and the S-matrix derived with the formulation
of Ref.~\cite{ger:80}. 
These results, as well as additional properties of the inverse square potential
and other singular interactions, will be presented elsewhere.

\section{Quantum Field Theory}
\label{sec:QFT}

Some comments on the quantum field theory case are now in order. The
connection to this work is twofold:

(i) The emphasis on singular potentials 
that we have adopted in this paper can be traced
fundamentally to their field-theory origin.
This explicit connection is illustrated below, in the final paragraph of this section.

(ii) The approach initiated in this paper provides the underlying philosophy
and hopefully a partial set of ingredients for the more general program of bound states in 
quantum field theory.
This generalization is currently in progress.

In particular, the definitions of Sec.~\ref{sec:QM_propagator}
can be adjusted as follows.
Let us consider the path integral 
\begin{equation} 
\left\langle 
\phi_{2} ({\bf x}), t_{2} \left| \right. 
\phi_{1}({\bf x}), t_{1}
\right\rangle   
 = 
\int_{\phi( {\bf x}, t_{1})  
=  \phi_{1} ( {\bf x}) }^{ \phi({\bf x} , t_{2}) 
 =  \phi_{2} ( {\bf x} ) }  
 \;  
{\cal D}  \phi  \,
\exp \left\{ {\frac{i}{\hbar} S [\phi]} \right\}
\;  
\label{eq:position_propagator}
\end{equation}
of the field $\phi ({\bf x})$ between the states 
$\left. \left. \right| \phi_{1}({\bf x}, t_{1}) \right\rangle $
and
$\left. \left. \right| \phi_{2}({\bf x}, t_{2}) \right\rangle $, 
corresponding to the propagator 
\begin{equation} 
\left\langle 
\phi_{2} ({\bf x}), t_{2} \left| \right. 
\phi_{1}({\bf x}), t_{1}
\right\rangle 
=
\left\langle \phi_{2}  ({\bf x}) \left| 
\hat{T}
\exp \left[
 -\frac{i}{\hbar} \int_{t_{1}}^{t_{2}}
\hat{H} dt 
\right]
\right| \phi_{1}  ({\bf x}) \right\rangle   
\; ,
\label{eq:position_propagator2}
\end{equation}
with $\hat{T}$ being again the time-ordering operator
and  $\hat{H}$ the Hamiltonian.
For time-independent  Hamiltonians
the ``energy propagator'' is defined to be
\begin{eqnarray}
\mbox{\boldmath\Large  $\left(  \right.$ } \! \! \!  
\phi_{2}({\bf x}) 
\left| \right.
 \phi_{1}({\bf x}) 
\mbox{\boldmath\Large  $\left.  \right)$ } \! \! \!_{E}
&  = & 
\frac{1}{i \hbar}
\int_{t_{1}}^{\infty}
d t_{2}
\,
\left\langle \phi_{2} ({\bf x}), t_{2} | \phi_{1}({\bf x}), t_{1}
\right\rangle   
e^{i  E (t_{2}-t_{1})/\hbar }
\nonumber 
\\
& = &
\left\langle \phi_{2}({\bf x}) 
\left|
\left( 
E - \hat{H} + i \epsilon 
\right)^{-1}
 \right|
\phi_{1}({\bf x}) \right\rangle   
\; 
\label{eq:energy_propagator}
\end{eqnarray}
(with  $i\epsilon$ being a small positive imaginary part),
which admits a spectral representation
\begin{widetext}
\begin{equation}
\mbox{\boldmath\Large  $\left(  \right.$ } \! \! \!  
\phi_{2}({\bf x}) 
\left| \right.
 \phi_{1}({\bf x}) 
\mbox{\boldmath\Large  $\left.  \right)$ } \! \! \!_{E}
=
\sum_{n}
\frac{\Psi_{n}[\phi_{2}({\bf x})]  \,
\Psi_{n}^{*}[\phi_{1}({\bf x})]}{E-E_{n} +i \epsilon}
+
\int d \alpha
\,
\frac{\Psi_{\alpha}[\phi_{2}({\bf x})]  \,
\Psi_{\alpha}^{*}
[\phi_{1}({\bf x})]}{E-E_{\alpha} +i \epsilon}
\;  
\label{eq:energy_propagator_states_QFT}
\end{equation}
\end{widetext}
and permits, in principle, the identification
of bound states and scattering states, as well as the corresponding
wave functionals.

A subtlety arises in
the usual description of bound states at low energies, which requires a
nonrelativistic (nr) reduction to be performed in 
one of the following ways:
(i) application of the nr limit
to the original path integral~(\ref{eq:position_propagator}),
followed by evaluation of the Fourier transform, to be
subsequently expanded as in~(\ref{eq:energy_propagator_states_QFT});
(ii) application of the nr limit
at the level of Eq.~(\ref{eq:energy_propagator_states_QFT}).
It is not obvious that procedures (i) and (ii) are equivalent.
In nuclear physics at low energies one may wish
to integrate out the light field 
(e.g., pions) and describe the dynamics in terms of the heavy field only (i.e.,
nucleons). Issues about the order in which to take the nr limit are also
present in this case. 
Regardless of the route taken, the appearance of
singular potentials is a generic feature of these nr reductions from field
theory to quantum mechanics~\cite{Stevenson,Beane}. 
The proper treatment
of these potentials is notoriously difficult and requires regularization and
renormalization of the corresponding  propagator.
This particular problem has been addressed for the inverse square potential in
Sec.~\ref{sec:ISP_regularization}.

An intriguing application of these ideas can be found in the recent 
literature~\cite{car:01}
 in relation to the study
of bound states for the Yukawa model,
\begin{equation}
{\cal L}_{\rm int}
=
g \bar{\Psi} \Psi 
\varphi
\;  ,
\end{equation}
where $\Psi$ represents
fermions and $\varphi$ the scalar interaction. 
For two-fermion wave functions of total angular momentum
$ J= 0^{+}$,
and within a covariantized light-cone
technique,  the stability of  bound states can be assessed
by the asymptotics of an effective wave equation that  simulates a
nonrelativistic inverse square potential.
Additional subtleties 
of this problem arising from the renormalized strong-coupling
sector of the theory are currently under investigation.

\section{Conclusions}
\label{sec:conclusions}

In summary, we have formulated the general outline of a program for the evaluation 
of quantum-mechanical propagators and energy Green's functions.
We have also thoroughly examined the inverse square potential
within such path-integral framework.
Our analysis included renormalization {\it \`{a} la\/} field theory
in the (strong) supercritical regime, as well as the inclusion of an additional
delta-function interaction.
The general technique as well as the particular examples
suggest that: 

(i)  The treatment of singular potentials---and of bound states in 
general---is most
efficiently accomplished by the use of 
the Fourier-transformed or energy Green's function $G(E)$.

(ii) Infinite summations and resummations in the spirit of perturbation theory
capture
the required nonperturbative behavior associated with the bound-state
sector of the theory.
 
(iii) Proper analytic continuations are needed in certain regimes.

(iv) The effective-field-theory program, which  
leads to singular potentials, requires renormalization
in a quantum-mechanical setting---renormalization that,
in principle, could be implemented with
techniques similar to the ones presented in this paper.

Extensions of this generic program will be presented elsewhere.

\begin{acknowledgments}
This research was supported in part by
an Advanced Research Grant from the Texas
Higher Education Coordinating Board 
and by the University of San Francisco Faculty Development Fund.
Stimulating
discussions with Profs.\ 
Luis N. Epele, Huner Fanchiotti,
 and Carlos A. Garc\'{\i}a Canal
are gratefully acknowledged by H.E.C.
We also thank Prof.\ D. K. Park
for bringing Refs.~\cite{Randall-Sundrum,Park} to our attention.

\end{acknowledgments}

\appendix*

\section{Formal Operator Structure of the Perturbation Expansion}
\label{sec:formal_operator_expansion}

In this appendix we summarize the formal algebraic structure of the
operator expansions presented in Sec.~\ref{sec:integral_representations}.
In addition,
we review the necessary ingredients of the Kontorovich-Lebedev transform, 
properly adjusted to the notation and goals of this paper,
and with emphasis on the results of 
Secs.~\ref{sec:integral_representations} and \ref{sec:integral_representation_ISP}.

A normalized continuous basis
for the space of square integrable functions over the interval
${\mathcal I} \in 
\mathbb{R} $
is provided by the set of
functions $\left\{ \Psi_{z} (q) \right\}_{z \in {\mathcal J}}$
that satisfy the orthonormality condition
\begin{equation}
\int_{\mathcal I}
dq
\,
\Psi_{z''}^{*}(q) \, \Psi_{z'} (q)
=\delta (z''-z')
\;
\label{eq:orthonormality}
\end{equation}
and
the completeness relation
\begin{equation}
\int_{\mathcal J}
dz 
\,
\Psi_{z} (q) \, \Psi_{z}^{*} (q')
= \delta (q-q')
\;  ,
\label{eq:completeness}
\end{equation} 
with the continuous
variable $z$ defined over an interval
${\mathcal J} \in  \mathbb{R} $.
A particular set
$\left\{ \Psi_{z} (q) \right\}_{z \in {\mathcal J}}$ can be constructed
by selecting the eigenbasis associated with an appropriate self-adjoint operator,
with properly chosen boundary conditions.
The mathematical theory required for such bases, which is extremely subtle,
should be studied carefully for every specific case,
and no attempt is made to develop a rigorous presentation in this paper.
Within that operational 
framework, we adopt the usual 
Dirac notation, which reproduces 
the required formal algebraic structure of
Eqs.~(\ref{eq:orthonormality}) 
and (\ref{eq:completeness}) from
\begin{equation}
\left\langle 
\Psi_{z''} 
|
\Psi_{z'} 
\right\rangle
= \delta (z''-z')
\;
\label{eq:orthonormality_Dirac}
\end{equation}
and
\begin{equation}
\int_{\mathcal J}
dz 
\,
\left|
\Psi_{z} 
\right\rangle
\,
\left\langle 
\Psi_{z} 
\right|
=
\openone
\;  .
\label{eq:completeness_Dirac}
\end{equation} 
In particular, the  basis $\left\{ \left| q  \right\rangle \right\}$
provides direct contact with
the functions $\Psi_{z} (q) \equiv \left\langle q| \Psi_{z} \right\rangle$.
 These are the functions used in the expansions of
 Sec.~\ref{sec:integral_representations},
where $q \equiv r$ represents the radial variable.

The most important results of Secs.~\ref{sec:integral_representations} and 
\ref{sec:integral_representation_ISP}
rely on the properties of operators. With that purpose in mind, an
operator $\hat{A}$ is represented by the ``continuous matrix''
$A(q'',q') =
\left\langle
 q'' |A| q'
\right\rangle
$  in the basis $\left\{ \left| q  \right\rangle \right\}$.
More generally, $\hat{A}$ admits the representation
$\left\langle
\Psi_{z''}
 |A| 
\Psi_{z'}
\right\rangle
$
in the
$\left\{  \left| \Psi_{z} \right\rangle
 \right\}_{z \in {\mathcal J}}$,
 in terms of which the sesquilinear expansion 
\begin{equation}
\left\langle 
q'' |A| q'
\right\rangle
=
\int_{\mathcal J}
dz'' 
\int_{\mathcal J}
dz' 
\,
\Psi_{z''} (q'')
\,
\left\langle
\Psi_{z''}
 |A| 
\Psi_{z'}
\right\rangle
\,
\Psi_{z'}^{*}(q') 
\;
\label{eq:sesquilinear}
\end{equation}
directly follows
by double insertion of the completeness identity~(\ref{eq:completeness_Dirac}).

Equation~(\ref{eq:sesquilinear}) is further simplified for the particular case in which 
$\left| 
\Psi_{z'}
\right\rangle
$
represents the eigenbasis of a self-adjoint operator $\hat{A}$,
with eigenvalues $a(z)$. In that case,
\begin{equation}
\left\langle 
q'' |A| q'
\right\rangle
=
\int_{\mathcal J}
dz
\,
a(z)
\,
\Psi_{z} (q'')
\,
\Psi_{z'}^{*}(q') 
\;
\label{eq:spectral_reps}
\end{equation}
is the required spectral representation of the operator,
which we have directly applied, for example, to 
Eqs.~(\ref{eq:GreenF_spectral_reps}),
(\ref{eq:GreenF_order_zero_spectral_reps}),
and (\ref{eq:GreenF_KL}).

The integral  transform
of the function $f(q)$ by the kernel
$K(z,q)$ is defined by
\begin{equation}
{\mathcal T} 
\left\{
f(q); K(z,q)
\right\} (z)
\equiv 
\tilde{f}
=
\int_{\mathcal I} 
dq 
\,
K(z,q) f(q)
\;  ,
\end{equation}
 with appropriate 
conditions required for the convergence of the integrals
involved. 
Of particular interest for the derivations of Secs.~\ref{sec:integral_representations} and 
\ref{sec:integral_representation_ISP}
is the class of integral transforms
that arise from generalized bases;
specifically,
given the basis
$\left\{  \left| \Psi_{z} \right\rangle
 \right\}_{z \in {\mathcal J}}$,
the associated transform is defined by  
\begin{equation}
{\mathcal T} 
\left\{
f(q); K(z,q)
\right\} (z)
=
\int_{\mathcal J}
dq \,
\Psi_{z}^{*} (q) \,
f(q)
\;  ,
\label{eq:transform}  
\end{equation}
which amounts to
$K(z,q) \equiv \Psi_{z}^{*} (q)$.
Equation~(\ref{eq:transform} ) 
is just a particular set of ``components'' of the vector
$\left| f \right\rangle$, as can be seen with the Dirac 
notation
\begin{equation}
{\mathcal T} 
\left\{
f(q); K(z,q)
\right\} (z)
=
\left\langle \Psi_{z} | f \right\rangle 
\;   ,
\label{eq:generalized_transform}
\end{equation}
by insertion of the completeness relation for the $\left| q \right\rangle$ basis.
The 
abbreviated notation $\tilde{f}(z)=\left\langle z | f \right\rangle $ 
highlights that this class of integral transforms amounts to a mere change of basis in Hilbert
space---the analogue of the ``passive'' view of a coordinate transformation.
Reciprocally,
the integral transform~(\ref{eq:transform}) can be used to reproduce 
the original function by means of an inversion identity
\begin{equation}
f(q) =
\int_{\mathcal J}
 dz 
\,
\Psi_{z} (q)
 \,
\tilde{f}(z)
\;  ,
\label{eq:generalized_reciprocal_transform}
\end{equation}
which follows by 
insertion of the completeness relation for the generalized basis~(\ref{eq:completeness}).
A few of the most common examples of this
kind of integral transform are provided by the following partial list 
of transformation kernels or bases (in 1D):
(i) exponential Fourier transforms, $\Psi_{z}(q)= (2\pi)^{-1/2} e^{\pm iqz}$;
(ii) trigonometric Fourier transforms, $\Psi_{z}(q) = \sqrt{2/\pi}
\, \tau (qz)$,
with $\tau = \sin$ or $\tau = \cos$;
(iii)
Hankel (Fourier-Bessel) transforms,
$\Psi_{z}(q) = (qz)^{1/2} J_{\mu}(qz)$; etc.
However, this general formal structure is of interest
to gain familiarity with more exotic cases, such as the
Kontorovich-Lebedev transform, which
 is central to our analysis of the inverse square potential in 
Sec~.\ref{sec:integral_representation_ISP}.

The Kontorovich-Lebedev representation~\cite{kontorovich},
is based on the modified Bessel function
of the second kind with imaginary order
$K_{iz}(q)$, with the 
the relevant interval being the real half-line
${\mathcal I} = [0,\infty)$.
The notation
$q \equiv \kappa r$
proves convenient  for direct comparison with the equations of 
Sec.~\ref{sec:integral_representation_ISP}; it is noteworthy that, due to a simple
scaling argument (equivalent to scale
invariance of the inverse square potential), 
the corresponding Dirac-normalized complete set of functions
is
\begin{equation}
\Psi_{z} (r;\kappa)
=
\sqrt{  \frac{ 2 z \sinh ( \pi z )  }{ \pi^{2} r }  }
\,
K_{iz} (\kappa r) 
\;   .
\end{equation}
An important property of the Bessel functions $K_{iz}(q)$ for real variables
$q$ and $z$ is that
$
K_{iz}^{*}(q) =
K_{iz}(q)$,
i.e., it is real.
The orthonormality relation~(\ref{eq:orthonormality})
of the functions $\{ \Psi_{z} (r) \}_{ z \in {\mathcal J}}$, with ${\mathcal J} =[0,\infty)$,
amounts to the integral identity
\begin{equation}
\int_{0}^{\infty}
\frac{dr}{r}
\,
K_{iz''} (\kappa r) 
K_{iz'} (\kappa r) 
=
\frac{ \pi^{2}  }{ 2 z \sinh ( \pi z )  }
\,
\delta (z''-z')
\;  ,
\label{eq:orthonormality_KL}
\end{equation}
which can be verified
from the integral of the expression~\cite{wat:44b}
\begin{equation}
K_{iz''} (x) 
K_{iz'} (x) 
=
2 
\int_{0}^{\infty}
dt 
\,
K_{i(z'+z'')} (2x \cosh t) 
\cos \left[ 
\left( z''-z'
\right) t
\right]
\;  ,
\label{eq:orthonormality_KL_aux}
\end{equation}
with respect to $x$,
followed by an exchange of the order of integration
and explicit integration of the remaining expression~\cite{gra:00},
with an appropriate analytic continuation. 
Finally,
completeness is just equivalent to the Kontorovich-Lebedev inversion 
identity, 
\begin{equation}
\frac{2}{ \pi^{2}}
\int_{0}^{\infty} 
dz
\,
 z \sinh ( \pi z )  K_{iz} (x) dz
\,
\int_{0}^{\infty} \frac{dy}{  y }  
  K_{iz} (y)
f(y)
=
f  ( x)
\;
\end{equation}
(see Ref.~\cite{kontorovich})
for functions $f(x)$ defined on the positive real half-line with appropriate 
conditions required for convergence.


\begin{thebibliography}{99}

\bibitem{fra:71} 
W. M. Frank, D. J. Land, and R. M. Spector,
Rev. Mod. Phys. {\bf 43}, 36 (1971); and references therein.

\bibitem{wei:95} 
S. Weinberg,
 {\em The Quantum Theory of Fields\/}
(Cambridge University Press, Cambridge, 1995), Vol.\ I, p.\ 560. 

\bibitem{jac:private}
R. Jackiw, private communication;
F. Gross, private communication.

\bibitem{bet:51}
H. A. Bethe and E. E. Salpeter,
Phys. Rev. {\bf 82},  309 (1951);
{\bf 84},  1232 (1951).

\bibitem{recent_review}
G. P. Lepage, nucl-th/9706029;
U. van Kolck, Prog. Part. Nucl. Phys. {\bf 43}, 337 (1999);
S. R. Beane, P. F. Bedaque, M. J. Savage, and U. van Kolck,
nucl-th/0104030.

\bibitem{kleinert}
H. Kleinert, {\em Path Integrals in Quantum Mechanics, Statistics,
and Polymer Physics\/}, 2nd ed.\  
(World Scientific, Singapore, 1995);
and references therein.

\bibitem{sav:99}
For a recent approach based
on the Feynman-Schwinger representation, see
\c{C}. \c{S}avkli, J. Tjon, and F. Gross,
Phys. Rev. C {\bf 60}, 055210 (1999); {\bf 61}, 069901(E) (2000).

\bibitem{mot:49}
N. F. Mott and H. S. W. Massey,
{\em The Theory of Atomic Collisions\/},
2nd ed. (Oxford Univ. Press, Oxford, 1949),
p.\ 40;
K. M. Case,
Phys. Rev. {\bf 80}, 797 (1950);
L. D. Landau and E. M. Lifshitz,
{\em Quantum Mechanics\/}, 3rd ed.
(Pergamon, Oxford, 1977),
p.\ 114;
P. M. Morse and H. Feshbach,
{\em Methods of Theoretical Physics\/}
(McGraw-Hill, New York, 1953),
 Vol.\ 2, p.\ 1665.

\bibitem{gup:93}
K. S. Gupta and S. G. Rajeev,
Phys. Rev. D {\bf 48}, 5940 (1993). 

\bibitem{cam:00}
H. E. Camblong, L. N. Epele, H. Fanchiotti, and C. A. Garc\'{\i}a Canal, 
Phys. Rev. Lett. {\bf 85}, 1590 (2000).

\bibitem{cam:01}
H. E. Camblong, L. N. Epele, H. Fanchiotti, and C. A. Garc\'{\i}a
Canal,
Ann. Phys. (NY) {\bf 287}, 14 (2001);  57 (2001);
and references therein.

\bibitem{nel:64}
E. Nelson,
J. Math. Phys. {\bf 5}, 332 (1964).  

\bibitem{pea:69}
D. Peak and A. Inomata, 
J. Math. Phys. {\bf 10}, 1422 (1969).  

\bibitem{jar:88}
T. Jaroszewicz, Phys. Rev. Lett.
{\bf 61}, 2401 (1988).

\bibitem{bha:89}
K. V. Bhagwat and S. V. Lawande,
Phys. Lett. A {\bf 141}, 321 (1989).

\bibitem{gro:98} 
C. Grosche and F. Steiner,  
{\it Handbook of Feynman Path Integrals\/}
(Springer-Verlag, Berlin, 1998);
and references therein.

\bibitem{cam:01b}
H. E. Camblong, L. N. Epele, H. Fanchiotti, and C. A. Garc\'{\i}a
Canal, 
 Phys. Rev. Lett. {\bf 87}, 220402  (2001).

\bibitem{extra_terms_qm}
B. DeWitt, Rev. Mod. Phys.
{\bf 29} (1957) 377;
S. F. Edwards and Y. V. Gulyaev, 
Proc. R. Soc. A {\bf 279}, 229 (1964);
J.-L. Gervais and  A. Jevicki, 
Nucl. Phys. B {\bf 110}, 93 (1976); 
K. M. Apfeldorf
 and C. Ord\'o\~nez, 
Nucl. Phys. B {\bf 479}, 515 (1996);
and references therein.

\bibitem{extra_terms_qft}
K. M. Apfeldorf,
H. E. Camblong, and C. R. Ord\'{o}\~{n}ez,
Mod. Phys. Lett. {\bf 16}, 103 (2001).

\bibitem{erd:53}
A. Erd\'{e}lyi, W. Magnus, F. Oberhettinger, and F. G.
 Tricomi, eds.,
{\em Higher Transcendental Functions\/}
(McGraw-Hill, New York, 1955), Vol. 2, Chap. XI.

\bibitem{wat:44}
G. N. Watson, {\em A Treatise on the Theory of Bessel Functions\/},
2nd ed.\
(Cambridge University Press, Cambridge, England, 1944).

\bibitem{interdimensional}
J. H. Van Vleck, in {\em Wave Mechanics, the First Fifty Years\/},
W. C. Price {\em et al.\/}, eds.
(Butterworth, London, 1973), p.\ 26.

\bibitem{weber}
Equation~(\ref{eq:propagator_free_particle})
follows from repeated application of Weber's second exponential integral
(Ref.~\cite{wat:44}), as shown in Ref.~\cite{pea:69}. 

\bibitem{gra:00} 
I. S.  Gradshteyn and I. M. Ryzhik,
{\em Table of Integrals, Series, and Products\/},  6th ed.\
(Academic Press, New York, 2000), 
p.\ 705  (formula 6.653.1).

\bibitem{bha:88}
S. V. Lawande and K. V. Bhagwat,
Phys. Lett. A {\bf 131}, 8 (1988);
K. V. Bhagwat and S. V. Lawande,
Phys. Lett. A {\bf 135}, 417 (1988).

\bibitem{kontorovich}
A. Erd\'{e}lyi, W. Magnus, F. Oberhettinger, and F. G.
 Tricomi, eds.,
{\em Tables of Integral Transforms\/}
(McGraw-Hill, New York, 1954), Vol.\ 2, Chap.\ XII;
F. Oberhettinger,
{\em Tables of Bessel Transforms\/}
(Springer-Verlag, Berlin, 1972),
Chap.\ V;
S. B. Yakubovich,
{\em Index Transforms\/}
(World Scientific, Singapore, 1996).

\bibitem{gro:93}
C. Grosche, Phys. Rev. Lett. {\bf 71}, 1 (1993).

\bibitem{Randall-Sundrum}
L. Randall and R. Sundrum, 
 Phys. Rev. Lett. {\bf 83},  4690 (1999).

\bibitem{Park}
D. K. 
Park, hep-th/0108068.

\bibitem{delta}
C. Thorn,
Phys. Rev. D {\bf 19}, 639 (1979);
K. Huang,
{\em Quarks, Leptons, and Gauge Fields\/}
(World Scientific, Singapore, 1982), Secs.\ 10.7 and 10.8;
R. Jackiw,
in {\em M. A. B. B\'{e}g Memorial Volume\/},
A. Ali and P. Hoodbhoy, eds. (World Scientific, Singapore, 1991);
C. Manuel and R. Tarrach, 
  Phys. Lett. B {\bf 328}, 113 (1994); 
S. K.  Adhikari and T. Frederico, 
 Phys. Rev. Lett. {\bf 74}, 4572  (1995); 
and references therein.

\bibitem{cam:01c}
H. E. Camblong and C. R. Ord\'{o}\~{n}ez,
hep-th/0110176.

\bibitem{Green_operator_approach}
H. E. Camblong and C. R. Ord\'{o}\~{n}ez,
hep-th/0110278.

\bibitem{abr:72} 
M. Abramowitz and I. A. Stegun, eds.,
{\em Handbook of Mathematical Functions\/}
(Dover Publications, New York, 1972).

\bibitem{running_coupling}
Since we are working within the framework of effective Lagrangians, it
is natural to assume a cutoff ($a$) dependence of the ``bare'' 
coupling constant $\lambda$.

\bibitem{jac:72}
R. Jackiw, Physics Today {\bf 25}, No. 1, 23 (1972).

\bibitem{alf:76}
V. de Alfaro, S. Fubini, and G. Furlan, Nuovo Cimento A {\bf 34},
569 (1976).

\bibitem{jac:80}
R. Jackiw, Ann. Phys. (N.Y.) {\bf 129}, 183 (1980);
R. Jackiw, Ann. Phys. (N.Y.) {\bf 201}, 83 (1990).

\bibitem{ger:80}
C. C. Gerry and V. A. Singh,
Phys. Rev. D {\bf 21}, 2979 (1980).

\bibitem{Stevenson}
M. Consoli and P. M. Stevenson,
hep-ph/9711449; Int. J. Mod. Phys. A {\bf 15}, 133 (2000).
These authors
consider the effect of a $1/r^{3}$ potential coming 
from radiative corrections.

\bibitem{Beane}
S. R. Beane {\em et al.\/},
quant-ph/0010073.

\bibitem{car:01}
M. Mangin-Brinet, J. Carbonell, and V. A. Karmanov,
hep-th/0107235;
V. A. Karmanov, J. Carbonell, and  M. Mangin-Brinet, 
contribution to {\em Mesons and Light Nuclei '01\/},
Pragua, July 2001,
hep-th/0107235.

\bibitem{wat:44b}
See Ref.~\cite{wat:44},
Eq.~(1) on 
p.\ 440.

\bibitem{gra:00b}
See Ref.~\cite{gra:00}, 
p.\ 668  (formula 6.561.16).
 
\end{thebibliography}
\end{document}